\documentclass[journal]{IEEEtran}
 
\usepackage{algorithm}
\usepackage{algpseudocode}

\usepackage{color}
\usepackage{graphicx, subfigure}
\usepackage{amsmath}
\usepackage{cite}
\usepackage{amsfonts}
\usepackage{multirow}
\usepackage{url}
\usepackage{threeparttable}
\usepackage{rotating}
\usepackage{hyperref}
\usepackage{booktabs}
\usepackage{makecell} 
\usepackage[many]{tcolorbox}

\usepackage{colortbl,xcolor}  
\definecolor{Gray}{gray}{0.92}

\begin{document}

\title{FPNet: Joint Wi-Fi Beamforming Matrix Feedback and Anomaly-Aware Indoor Positioning}

\author{Ran~Tao,~\IEEEmembership{Student Member,~IEEE,}
        Jiajia~Guo,~\IEEEmembership{Member,~IEEE,}
        Yiming~Cui,~\IEEEmembership{Student Member,~IEEE,}
        Xiangyi~Li,~\IEEEmembership{Student Member,~IEEE,}
        Chao-Kai~Wen,~\IEEEmembership{Fellow,~IEEE,}
        and~Shi~Jin,~\IEEEmembership{Fellow,~IEEE}
\thanks{An earlier version of this work was presented in IEEE ICC \cite{9839120}.}

}

\maketitle

\begin{abstract}
Channel State Information (CSI) provides a detailed description of the wireless channel and has been widely adopted for Wi-Fi sensing, particularly for high-precision indoor positioning. However, complete CSI is rarely available in real-world deployments due to hardware constraints and the high communication overhead required for feedback. Moreover, existing positioning models lack mechanisms to detect when users move outside their trained regions, leading to unreliable estimates in dynamic environments.
In this paper, we present FPNet, a unified deep learning framework that jointly addresses channel feedback compression, accurate indoor positioning, and robust anomaly detection (AD). FPNet leverages the beamforming feedback matrix (BFM), a compressed CSI representation natively supported by IEEE 802.11ac/ax/be protocols, to minimize feedback overhead while preserving critical positioning features. To enhance reliability, we integrate ADBlock, a lightweight AD module trained on normal BFM samples, which identifies out-of-distribution scenarios when users exit predefined spatial regions.
Experimental results using standard 2.4 GHz Wi-Fi hardware show that FPNet achieves positioning accuracy above 97\% with only 100 feedback bits, boosts net throughput by up to 22.92\%, and attains AD accuracy over 99\% with a false alarm rate below 1.5\%. These results demonstrate FPNet's ability to deliver efficient, accurate, and reliable indoor positioning on commodity Wi-Fi devices.

\end{abstract}


\section{Introduction}
 
\IEEEPARstart{I}{ntegrated} sensing and communication (ISAC) is emerging as a significant trend for future wireless communication systems~\cite{singh2025integrated, zhu2024enabling,10926861}. Wi-Fi communication has become indispensable in modern society due to its high data rates and ease of deployment for wireless internet access. This widespread adoption has enabled Wi-Fi to support sensing applications such as indoor positioning, gesture recognition, and fall detection~\cite{du2024overview, meneghello2022sharp, meneghello2023toward, miao2025wi}, enhancing the functionality and versatility of Wi-Fi networks. The IEEE 802.11 working group established a dedicated Task Group, IEEE 802.11bf, to develop an amendment to the wireless local area network (WLAN) standard specifically targeting advanced sensing requirements while minimizing interference with existing communication functionalities~\cite{du2024overview}. Among various WLAN sensing tasks, indoor positioning is a critical enabling technology, providing fundamental support for many other sensing and communication applications~\cite{farahsari2022survey}.

In traditional indoor positioning systems, the Received Signal Strength Indicator (RSSI) has been widely adopted due to its simplicity and low implementation cost~\cite{ma2019wifi}. However, RSSI-based methods often suffer from limited accuracy, high susceptibility to interference, and significant performance degradation caused by multipath effects~\cite{heurtefeux2012rssi}. These limitations have motivated research into more precise and robust alternative techniques, among which Channel State Information (CSI) has emerged as a superior candidate.
 
CSI provides a more detailed description of the wireless channel compared to RSSI, capturing amplitude and phase information across multiple subcarriers and thereby offering greater accuracy and robustness in complex indoor environments~\cite{yang2013rssi}. Indoor positioning tasks based on CSI typically fall into two categories: regression and classification. Regression methods directly predict the exact coordinates (e.g., the $x$ and $y$ positions) of the user, providing high precision at the cost of increased computational complexity and data demands. In contrast, classification methods discretize the indoor environment into predefined zones and classify the user's position into these zones, offering greater simplicity and robustness under uncertain and dynamic conditions. Recent advancements in artificial intelligence (AI) and deep learning (DL) have significantly enhanced both regression and classification approaches. In particular, deep neural networks (NNs) have proven effective at mapping complex, high-dimensional CSI data to physical locations, leading to substantial performance improvements in indoor positioning systems~\cite{xue2022enhanced}.

Numerous recent studies have demonstrated the effectiveness of CSI-based positioning techniques. For instance, the fine-grained indoor fingerprinting system (FIFS) utilizes weighted average CSI values across multiple antennas to achieve precise position identification~\cite{xiao2012fifs}. Similarly, PinLoc employs CSI data trained on fine-grained $1\times1\,\mathrm{m}^2$ grid spots, greatly enhancing positioning granularity~\cite{pinloc}. CSI-MIMO combines the $k$-nearest neighbors algorithm with multi-input multi-output CSI measurements to achieve accurate positioning while maintaining low computational complexity~\cite{ibrahim2020csi}. Hi-Loc introduces a hybrid positioning framework that exploits enhanced 5G New Radio (NR) CSI by effectively integrating fingerprinting and geometric methods, thereby achieving high accuracy and robustness in challenging indoor scenarios~\cite{li2022hiloc}. CRISLoc proposes a reconstructable CSI fingerprinting scheme specifically designed for indoor smartphone positioning, leveraging transfer learning to adapt efficiently to environmental changes and achieve superior positioning accuracy~\cite{gao2019crisloc}. Furthermore, attention mechanisms within deep NNs, for example, attention-augmented residual convolutional networks, have been shown to significantly enhance feature extraction capabilities, thus improving positioning robustness and accuracy~\cite{liu2023csi}. These studies collectively underscore the immense potential of CSI as a foundational technology for high-precision indoor positioning.

Despite extensive work on CSI-based positioning, existing methods exhibit two key limitations.
First, although CSI is crucial for positioning, most current research assumes access to complete CSI, which poses practical challenges. To enable CSI-based sensing, researchers have introduced custom firmware and driver modifications that allow extraction of fine-grained CSI measurements. However, this functionality is supported by only a limited set of commercial Wi-Fi devices, such as the Intel 5300~\cite{CSItool}, Atheros chipsets~\cite{xie2015precise}, and other platforms enhanced through open-source tools~\cite{gringoli2019free,jiang2021eliminating}. These approaches, while effective, increase system complexity and limit scalability on commodity hardware.
Only about 6\% of more than 38,000 off-the-shelf Wi-Fi devices have been reported to support CSI extraction~\cite{wu2023enabling}, which has impeded its widespread deployment in practice. The growing adoption of IEEE 802.11ax, however, offers new opportunities as multi-user multiple-input multiple-output (MU-MIMO) support becomes increasingly common in commercial hardware.
Specifically, existing Wi-Fi feedback protocols from IEEE 802.11ac to IEEE 802.11be provide the beamforming feedback matrix (BFM) as the only feedback directly accessible at the access point (AP)~\cite{ieee80211ax2021}.

The advantages of BFM over CSI in terms of accessibility and ubiquity give it substantial potential in sensing applications, attracting considerable interest among researchers. Enze Yi \textit{et al.}~\cite{yi2024bfmsense} propose a BFM-ratio model that enables fine-grained BFM sensing for the first time. Subsequent studies have further demonstrated that BFM achieves good performance in tasks such as object recognition, firmware-agnostic line-of-sight identification, and action recognition\cite{jiang2022design, huang2025isac, shimomura2024beamforming, wang2024wi2dmeasure}. However, all of these BFM-based sensing tasks assume the availability of \emph{perfect} BFM. In practice, the BFM undergoes compression during the feedback process to reduce communication overhead, potentially leading to performance degradation. DL-based techniques have been applied to minimize CSI/BFM feedback overhead~\cite{9931713}, starting with CsiNet~\cite{wen2018deep}, a DL-based feedback architecture inspired by image-compression autoencoders. LB-SciFi~\cite{LB-SciFi2020} compresses sets of angles from the BFM using autoencoders, while EFNet~\cite{qifan2024} leverages frequency-domain correlations. Existing DL-based CSI compression methods primarily emphasize communication efficiency and overlook the explicit preservation of location-specific information, potentially compromising positioning accuracy. New compression approaches have been proposed to alleviate this burden. For example, Guo \textit{et al.}~\cite{guo2024learning} introduce a joint optimization framework for CSI feedback and positioning; however, it was developed and evaluated on the DeepMIMO~\cite{alkhateeb2019deepmimo} dataset for mobile communications without demonstrating feasibility in practical Wi-Fi deployments. 

Second, when a user moves outside a predefined region, the positioning system continues to estimate the user's location based on a model trained under the assumption that the user remains within that area. This can result in inaccurate or unreliable estimates, as the model has never encountered such out-of-distribution (OOD) samples during training. In practical applications, this poses a significant problem, particularly in dynamic and unknown environments.  

To comprehensively address these challenges, we propose FPNet, a unified DL framework explicitly designed to jointly optimize BFM feedback compression, accurate indoor positioning, and robust anomaly detection (AD). FPNet employs an end-to-end architecture with a shared encoder that compresses BFM data into low-dimensional codewords. These codewords are optimized concurrently for efficient BFM reconstruction and precise user positioning, ensuring that critical positioning information is preserved during compression and significantly enhancing both positioning accuracy and communication efficiency. To improve robustness against OOD scenarios, we integrate a lightweight AD module, ADBlock, into FPNet. Trained exclusively on normal BFM samples, ADBlock efficiently identifies anomalous BFM data during deployment with minimal computational overhead, enabling FPNet to detect when users move beyond predefined coverage areas and prevent misleading positioning results.

The key contributions of this paper are summarized as follows:
\begin{itemize}
  \item \textbf{Unified BFM Compression and Positioning Framework:} We propose FPNet, a novel DL framework explicitly designed to address practical challenges in Wi-Fi sensing by simultaneously compressing the BFM and preserving critical positioning information. FPNet employs a shared encoder structure optimized to retain positioning-relevant features based on imperfect BFM, enabling accurate indoor positioning while adhering to existing Wi-Fi protocol constraints.
  \item \textbf{Integration of Lightweight AD (ADBlock):} We introduce ADBlock, an autoencoder-based AD module seamlessly integrated within the FPNet framework. Trained exclusively on normal BFM samples, ADBlock efficiently identifies OOD scenarios where users move outside predefined positioning areas, substantially enhancing system robustness and reliability with minimal computational complexity.
    \item \textbf{Extensive Experimental Validation in Real-World Environments:} 
    We conduct comprehensive evaluations of FPNet using datasets collected with standard Wi-Fi hardware at 2.4 GHz.
    Experimental results demonstrate that FPNet achieves positioning accuracy greater than 97\% with just 100 feedback bits, improves net throughput by up to 22.92\%, and attains AD accuracy exceeding 99\% with a false alarm rate below 1.5\%, significantly outperforming conventional methods.
\end{itemize}

An earlier version of this work was presented at IEEE ICC \cite{9839120}, where the foundational FPNet architecture for joint BFM feedback and indoor positioning was introduced. This journal extension addresses a key limitation of the original framework, namely its vulnerability to OOD scenarios in which users move beyond the trained regions. Unlike the conference version \cite{9839120}, which focused on static environments, this work introduces ADBlock as a dedicated anomaly detection module and provides a comprehensive robustness analysis under dynamic conditions. Furthermore, the experimental evaluation is extended to include OOD detection performance and FPGA-based hardware latency analysis, validating the system practicality for real-world deployment.

The remainder of this paper is organized as follows. Section II details the system model. Section III presents the proposed FPNet framework, including the integration of AD and the role of compressed BFM. Section IV describes the experimental setup and evaluation results, including dataset collection, performance metrics, and quantitative analysis, highlighting improvements in positioning accuracy and throughput and demonstrating the effectiveness of the AD module. Finally, Section V concludes the paper and outlines possible directions for future improvements.

\section{System model and Problem Formulation}   
This section begins by introducing the channel sounding procedure in Wi-Fi systems, followed by an overview of the explicit feedback methods defined in IEEE 802.11ac and subsequent Wi-Fi standards~\cite{ieee80211ax2021}. Next, we discuss DL-based CSI feedback and positioning mechanisms. Finally, we highlight the challenge of handling OOD CSI samples in practical positioning scenarios, motivating the introduction of AD techniques to enhance system robustness. 

\subsection{Wi-Fi Channel Sounding Procedure}
\begin{figure}[t]
    \centering
    \setlength{\abovecaptionskip}{2mm}
    \includegraphics[width=0.99\linewidth,trim=1 2 1 2, clip]{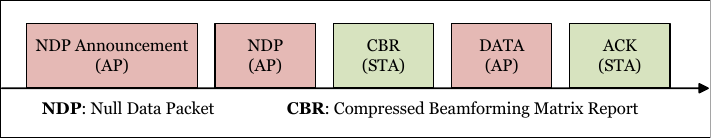}
    \caption{Channel sounding procedure in IEEE 802.11 Standard.}
    \label{feedback}
\end{figure}

In this paper, we consider a single-user MIMO (SU-MIMO) system in which the number of data streams is fewer than the number of transmit antennas, for simplicity. In a Wi-Fi SU-MIMO system, the AP and the client station (STA) are equipped with $N_\mathrm{t}$ and $N_\mathrm{r}$ antennas, respectively. As depicted in Fig.~\ref{feedback}, the channel sounding sequence begins with the AP transmitting an NDP Announcement (NDPA) frame, immediately followed by a Null Data Packet (NDP). The STA uses the received NDP for channel estimation, obtaining the CSI for each subcarrier and removing any cyclic shift delays. It then extracts the compressed beamforming matrix report (CBR) from the CSI and sends it back to the AP without encryption~\cite{yi2024bfmsense}. Once the AP receives this feedback, it constructs the precoding matrix for downlink transmission and applies it to positioning and other relevant tasks.

\subsection{Explicit Feedback in IEEE 802.11 Standard}

The received signal over the $k$-th subcarrier at the STA can be expressed as
\begin{equation}
\mathbf{y}[k] = \mathbf{H}[k]\mathbf{x}[k] + \mathbf{n}[k], \quad k = 0, 1, \ldots, N_\mathrm{c}-1,
\end{equation}
where $\mathbf{y}[k]\in\mathbb{C}^{N_\mathrm{r}\times 1}$, $\mathbf{H}[k]\in\mathbb{C}^{N_\mathrm{r}\times N_\mathrm{t}}$, $\mathbf{x}[k]\in\mathbb{C}^{N_\mathrm{t}\times 1}$, and $\mathbf{n}[k]\in\mathbb{C}^{N_\mathrm{r}\times 1}$ represent the received data packet, the channel matrix, the transmitted data packet, and the white Gaussian noise vector, respectively. Here, $N_\mathrm{c}$ denotes the total number of subcarriers in an orthogonal frequency-division multiplexing (OFDM) symbol. 

As the number of antennas and subcarriers in MIMO systems continues to grow, transmitting the entire  BFM  from the STA to the AP incurs significant overhead, reducing overall throughput. To mitigate this, the IEEE 802.11 standard employs a compression technique that reduces the BFM into two sets of angles. A detailed description is provided in~\cite{IEEE802.11ac}; a brief overview follows.

First, we apply singular value decomposition (SVD) to the channel matrix $\mathbf{H}[k]\in\mathbb{C}^{N_\mathrm{r}\times N_\mathrm{t}}$, yielding 
\begin{equation}
\mathbf{H}[k]=\mathbf{U}[k]\boldsymbol{\Sigma}[k]{\widetilde{\mathbf{V}}}^\mathrm{H}[k],
\label{eq:SVD}
\end{equation}
where $\mathbf{U}[k]\in\mathbb{C}^{N_\mathrm{r}\times N_\mathrm{r}}$ and $\widetilde{\mathbf{V}}[k]\in\mathbb{C}^{N_\mathrm{t}\times N_\mathrm{t}}$ are unitary matrices, and $\boldsymbol{\Sigma}[k]\in\mathbb{C}^{N_\mathrm{r}\times N_\mathrm{t}}$ is a diagonal matrix with nonnegative real entries. The beamforming feedback matrix $\mathbf{V}[k]$ is constructed by selecting the first $N_\mathrm{s}$ columns of $\widetilde{\mathbf{V}}[k]$, where $N_\mathrm{s}$ denotes the number of spatial streams configured for the STA. Subsequently, a series of Givens rotations parameterize $\mathbf{V}[k]$ using two sets of angles, $\mathbf{\Psi}$ and $\mathbf{\Phi}$: 
\begin{equation}
\mathbf{V}[k]=\left(\prod_{i=1}^{\min(N_\mathrm{s},N_\mathrm{t}-1)}\left(\mathbf{D}_i(\phi^{(k)})\prod_{l=i+1}^{N_\mathrm{t}-1}\mathbf{G}_{li}^\mathrm{T}(\psi_{li}^{(k)})\right)\right)\mathbf{I}_{N_\mathrm{t}\times N_\mathrm{s}}.
\end{equation}

In this formulation, $\mathbf{D}_i(\phi)$ represents an $N_\mathrm{t}\times N_\mathrm{t}$ diagonal matrix defined as
\begin{equation}
\left.\mathbf{D}_i(\phi)=\left[\begin{array}{ccccc}\mathbf{I}_{i-1}&0&0&\ldots&0\\0&e^{j\phi_{i,i}}&0&\ldots&0\\0&0&\ddots&0&0\\0&0&0&e^{j\phi_{N_\mathrm{t}-1,i}}&0\\0&0&0&0&1\end{array}\right.\right],
\end{equation}
and $\mathbf{G}_{li}(\psi)$ denotes an $N_\mathrm{t}\times N_\mathrm{t}$ Givens rotation matrix expressed as
\begin{equation}
\left.\mathbf{G}_{li}(\psi)=\left[\begin{array}{ccccc}\mathbf{I}_{i-1}&0&0&0&0\\0&\cos\psi&0&\sin\psi&0\\0&0&\mathbf{I}_{l-i-1}&0&0\\0&-\sin\psi&0&\cos\psi&0\\0&0&0&0&\mathbf{I}_{N_\mathrm{t}-l}\end{array}\right.\right],
\end{equation}
where $\mathbf{I}_m$ denotes an $m\times m$ identity matrix. Here, $\mathbf{I}_{N_\mathrm{t}\times N_\mathrm{s}}$ is a rectangular identity matrix (zero-padded if $N_\mathrm{t} \neq N_\mathrm{s}$) used to match the dimensionality.
Table~\ref{table:angles} summarizes the relationship between the size of the CSI feedback matrix and the corresponding number and order of Givens rotation angles ($\phi$ and $\psi$) required for compression. As the matrix size increases, the number of required angles grows accordingly, and their order follows the decomposition structure defined by the IEEE 802.11ac/ax standards.

\begin{table*}[t]
\centering
\caption{Numbers of $\phi$ and $\psi$ for 1 subcarrier in CSI Feedback Matrix~\cite{IEEE802.11ac}}
\label{table:angles}
\renewcommand{\arraystretch}{1.2}
\begin{tabular}{|c|c|l|c|c|}
\hline
\textbf{Size of matrix} & \textbf{Number of angles} & \multicolumn{1}{c|}{\textbf{The order of angles}} & \textbf{Type 0 Bits} & \textbf{Type 1 Bits} \\ \hline
$2 \times 1$ & 2 & $\phi_{11}, \psi_{21}$ & $7+5=12$ & $9+7=16$ \\ \hline
$2 \times 2$ & 2 & $\phi_{11}, \psi_{21}$ & $7+5=12$ & $9+7=16$ \\ \hline
$3 \times 1$ & 4 & $\phi_{11}, \phi_{21}, \psi_{21}, \psi_{31}$ & $7\times2 + 5\times2 = 24$ & $9\times2 + 7\times2 = 32$ \\ \hline
$3 \times 2$ & 6 & $\phi_{11}, \phi_{21}, \psi_{21}, \psi_{31}, \phi_{22}, \psi_{32}$ & $7\times3 + 5\times3 = 36$ & $9\times3 + 7\times3 = 48$ \\ \hline
$3 \times 3$ & 6 & $\phi_{11}, \phi_{21}, \psi_{21}, \psi_{31}, \phi_{22}, \psi_{32}$ & $7\times3 + 5\times3 = 36$ & $9\times3 + 7\times3 = 48$ \\ \hline
$\cdots$ & $\cdots$ & $\cdots$ & $\cdots$ & $\cdots$ \\ \hline
\end{tabular}
\end{table*}

To reduce feedback overhead, two additional steps are performed: adjacent‐carrier merging and angle quantization. Due to the high correlation among adjacent subcarriers, the STA can group them and feed back the BFM for only one subcarrier per group. According to IEEE 802.11ac, each group may contain 1, 2, or 4 subcarriers.

Since $\mathbf{\Psi}$ and $\mathbf{\Phi}$ have different compressibility, distinct quantization schemes are applied, with $b_{\psi}$ and $b_{\phi}$ denoting the number of bits allocated to $\mathbf{\Psi}$ and $\mathbf{\Phi}$, respectively~\cite{ieee80211ax2021}. Specifically, for Type 0 feedback, $b_{\psi}=5$ and $b_{\phi}=7$, while for Type 1, these values are 7 and 9. After the STA transmits the feedback, the AP reconstructs $\mathbf{V}[k]$ via inverse operations. The reconstructed $\mathbf{V}[k]$ matrices are then used for beamforming in subsequent transmissions. Although the standard’s feedback protocol achieves high accuracy, it incurs significant airtime overhead, potentially reducing actual data throughput.

\subsection{DL-based BFM Feedback and Positioning Mechanism}

In Wi-Fi systems, the STA must feed back the complete BFM, $\mathbf{V}$,  to the AP for beamforming and related tasks. Current BFM feedback mechanisms using autoencoders involve an encoder at the STA that compresses high-dimensional BFM matrices into low-dimensional codewords. These codewords are then quantized into bitstreams and sent over the uplink. At the AP, the feedback bits are processed through a dequantization module, followed by a decoder that reconstructs the original BFM from the compressed codewords. The entire process can be represented as
\begin{equation}
\widehat{\mathbf{V}} = f_{\text{de}}\left( \mathcal{D} \left( Q \left( f_{\text{en}} (\mathbf{V}) \right) \right) \right),
\end{equation}
where $f_{\text{en}} (\cdot)$ and $ Q \left(\cdot  \right)$ represent the compression and quantization operations at the STA, and $ \mathcal{D} \left(\cdot  \right)$ and $f_{\text{de}} (\cdot)$ denote the dequantization and reconstruction operations at the AP. The training loss function for optimizing this process is expressed as
\begin{equation}
    L_\mathrm{bfm} = \frac{1}{N}{\sum_{i=1}^{N} \| \widehat{\mathbf{V}}_i - \mathbf{V}_i \|_2^2} ,
\end{equation}
where $\widehat{\mathbf{V}}_{i}$ and $\mathbf{V}_{i}$ are the $i$-th reconstructed and original BFMs, respectively, and $N$ is the number of training samples. 

Because our dataset was collected in an indoor office environment, the primary positioning requirement is to determine the user's location within a specific area (e.g., identifying the office desk in use). We therefore simplify the positioning task as a classification problem, dividing the environment into discrete location spots, a methodology similar to that in~\cite{fukushima2019evaluating}. The dataset collection process is detailed in Section IV. 

The nonlinear mapping from the BFM to the STA’s position coordinates enables position estimation based on BFM data. For positioning, the NN takes the BFM as input and outputs a position category. We use the cross-entropy loss during training, defined as 
\begin{equation}
    L_\mathrm{pos} = - \frac{1}{N} \sum_{i=1}^N \sum_{j=1}^C y_{ij} \log(\hat{y}_{ij}) \label{eq:position},
\end{equation}
where $C$ is the total number of classes, $y_{ij}$ is the indicator (1 if sample $i$ belongs to class $j$, 0 otherwise), and $\hat{y}_{ij}$ is the predicted probability that sample $i$ belongs to class $j$.

\subsection{Autoencoder-Based Detection of Anomalous CSI Samples}
In Wi-Fi-based indoor positioning systems, a critical yet often overlooked challenge arises when a user moves into regions not covered during the training phase. Most machine learning models assume that input samples at inference time follow the same distribution as the training data. However, in practical deployments, this assumption is frequently violated due to dynamic environmental changes, unexpected user behavior, or the presence of users in previously unseen areas. This results in what is known as the OOD problem. 

To mitigate the risks associated with such inputs, AD techniques are commonly employed. The goal of AD is to determine whether a given input deviates significantly from the training distribution, allowing the system to flag or reject unreliable predictions. This capability is particularly important in safety-critical or high-reliability systems, where incorrect position estimates can have serious downstream consequences.

\begin{figure}[!t]
    \centering
    \includegraphics[width=1.0\linewidth, trim=120 90 140 140, clip]{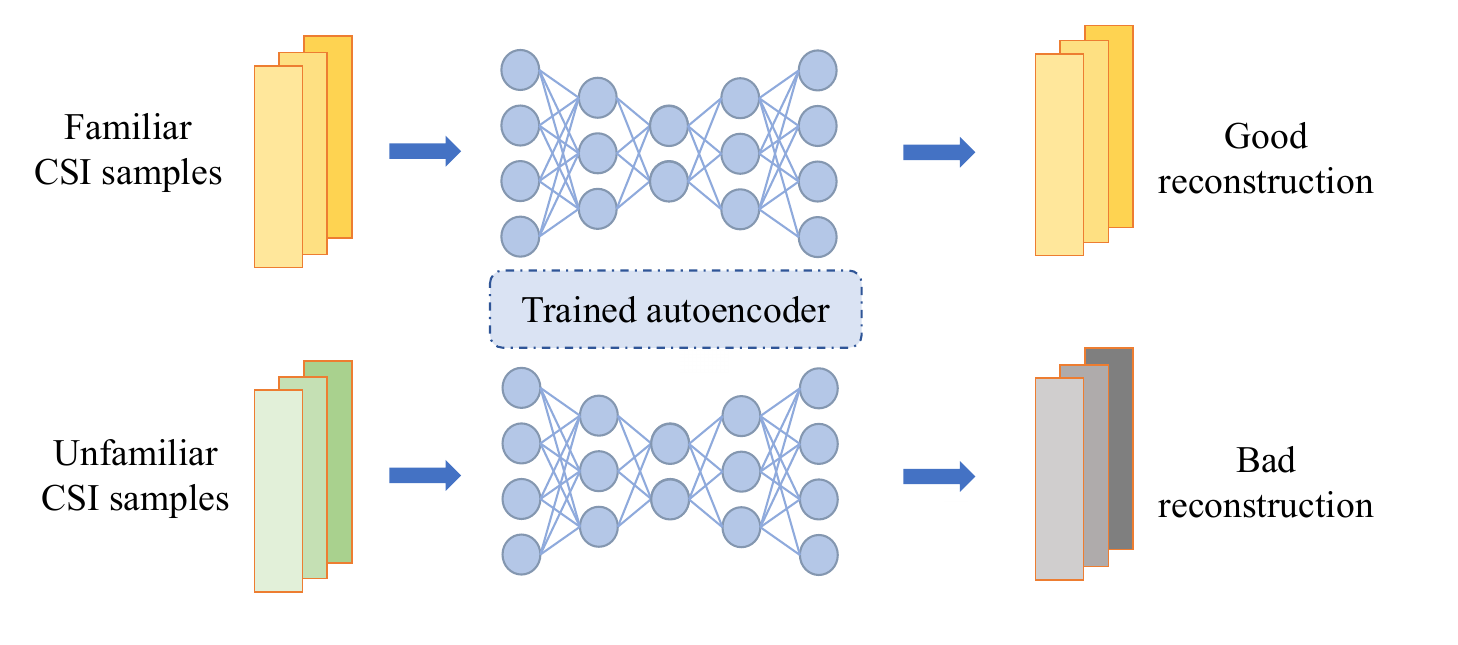}
    \caption{An illustration of reconstruction-based AD using a trained autoencoder. Familiar CSI samples (top) result in low reconstruction error, while unfamiliar samples (bottom) yield high reconstruction error and are thus flagged as anomalies.}
    \label{fig:ae_ad}
\end{figure}

A widely adopted approach is reconstruction-based AD using an autoencoder trained exclusively on normal data. During testing, inputs with high reconstruction errors are considered anomalous~\cite{zong2018deep, chalapathy2021deep}. Fig.~\ref{fig:ae_ad} illustrates this principle: familiar CSI samples (top row) are well reconstructed by the trained autoencoder, while unfamiliar (anomalous) CSI samples (bottom row) yield poor reconstructions and thus higher reconstruction errors. This discrepancy is used to flag anomalies in real time.
Other AD approaches include probabilistic modeling, latent feature distance measurement, and discriminative boundary-based methods such as One-Class SVM (OCSVM) and Isolation Forest (iForest)~\cite{khan2014survey}. In the context of CSI-based positioning, integrating an effective AD mechanism enhances robustness and deployment reliability. The specific implementation adopted in this work is detailed in Section~\ref{sec:architecture}.

\vspace{-0.6\baselineskip}
\section{FPNet}
\label{sec:architecture}

This section presents the overall design of FPNet, a unified DL framework that jointly addresses three critical tasks in Wi-Fi sensing: BFM feedback compression and reconstruction, indoor positioning, and AD. We first motivate the joint optimization of feedback and positioning, then describe the overall FPNet framework, next present the detailed AD mechanism, and finally outline the NN architectures for each component.

\subsection{Motivation}

\begin{figure}[t]
    \centering
    \includegraphics[width=0.9\linewidth]
    {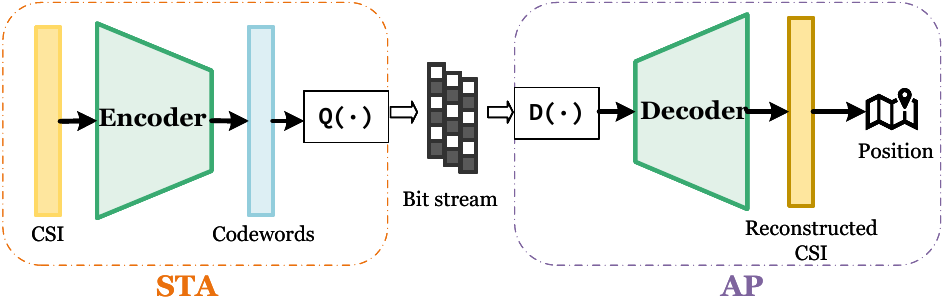}
    \caption{Current CSI-based positioning framework where the reconstructed CSI is used for positioning.}
    \label{sequential}
\end{figure}

Most existing CSI/BFM-based indoor positioning methods assume access to complete and accurate CSI for optimal performance. As shown in Fig.~\ref{sequential}, these methods typically extract CSI from compatible devices and feed it into a NN trained on labeled data. The network learns a mapping between high-dimensional CSI features and spatial positions, enabling precise predictions during inference.

However, in practical Wi-Fi systems compliant with IEEE 802.11 standards, CSI is transformed into a BFM, which is then compressed and quantized for uplink feedback to reduce communication overhead. This transformation often introduces distortions that degrade the fidelity of the CSI representation. Since accurate positioning critically depends on the quality of the BFM, such distortions can significantly impair performance. Existing feedback methods typically employ mean squared error (MSE) loss to optimize reconstruction quality without explicitly preserving spatial information relevant to positioning~\cite{yang2020deep, he2025cfnet}. To mitigate this issue, it is essential to retain positioning-relevant features during compression.

This principle mirrors that of image compression: two images with similar pixel-wise reconstruction quality can yield vastly different results when used for downstream tasks such as object recognition~\cite{ozah2019compression}. Therefore, task-aware compression, where compressed codewords are jointly optimized for feedback and positioning, is crucial. This motivates the design of a shared encoder structure that supports both BFM reconstruction and position estimation, as introduced in our FPNet framework.

In addition to the need for joint optimization, real-world deployment presents another critical challenge: the presence of previously unseen or OOD BFM samples. These occur when users enter regions outside the predefined spatial areas or when environmental changes significantly alter channel conditions. Traditional positioning models, which assume all inputs follow the training distribution, may produce unreliable or misleading estimates under such circumstances. 

To address this limitation, we integrate an unsupervised AD mechanism into our framework. By modeling the distribution of normal BFM patterns during training, the system can identify anomalous inputs at inference, those that deviate significantly from the expected data manifold, thus enhancing robustness and reducing the risk of incorrect positioning decisions. This combination of joint task optimization and anomaly-aware inference forms the core motivation behind FPNet.

\vspace{-0.6\baselineskip}
\subsection{FPNet Framework}

\begin{figure}[t]
    \centering
    \setlength{\abovecaptionskip}{2mm}
    \includegraphics[width=0.97\linewidth]
    {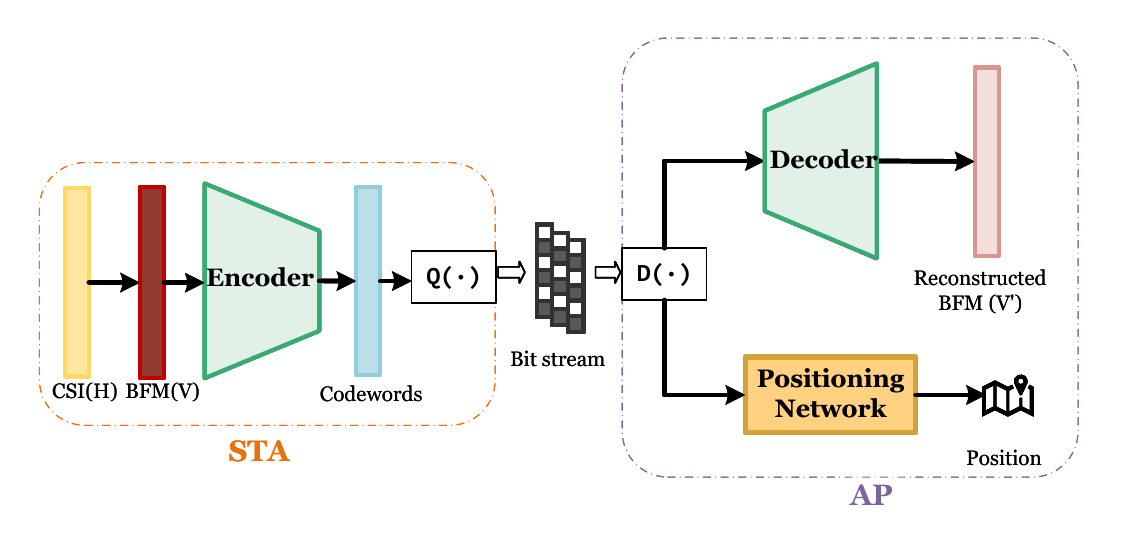}
    \caption{Illustration of the FPNet framework. The compressed BFM is used for both reconstruction and positioning.}
    \label{parallel}
\end{figure}

In this subsection, we describe the FPNet framework, focusing on the architecture and design for joint BFM feedback and positioning. While the overall framework also includes an AD module, its details are covered in the next subsection. Here, we concentrate on the components related to BFM compression and positioning. 
The FPNet framework consists of three primary components: an encoder for BFM compression, a decoder for BFM reconstruction, and a positioning network for estimating the user's position, as illustrated in Fig.~\ref{parallel}.

In the first step, SVD is applied to the CSI matrix $\mathbf{H}[k]$ for all valid subcarriers in each packet, yielding the set of matrices $\{\mathbf{V}[k]\}_{k=0}^{N_{\mathrm{vs}}-1}$, where $N_{\mathrm{vs}}$ is the number of valid subcarriers. The first $N_\mathrm{s}$ columns of each $\mathbf{V}[k]$ are extracted and concatenated to form the BFM $\mathbf{V} \in \mathbb{C}^{N_{\mathrm{vs}}\times (N_{\mathrm{t}}N_{\mathrm{s}})}$.
The encoder extracts essential features from the BFM across adjacent subcarriers and compresses them into low-dimensional codewords. These codewords are quantized into bitstreams and transmitted to the AP, which uses them for both BFM reconstruction and user positioning. FPNet thus operates as a multi-objective NN, leveraging DL frameworks' capability to handle multiple outputs concurrently.

\begin{figure*}[!t]
    \centering
    \includegraphics[width=0.98\linewidth, trim=0 165 0 160, clip]{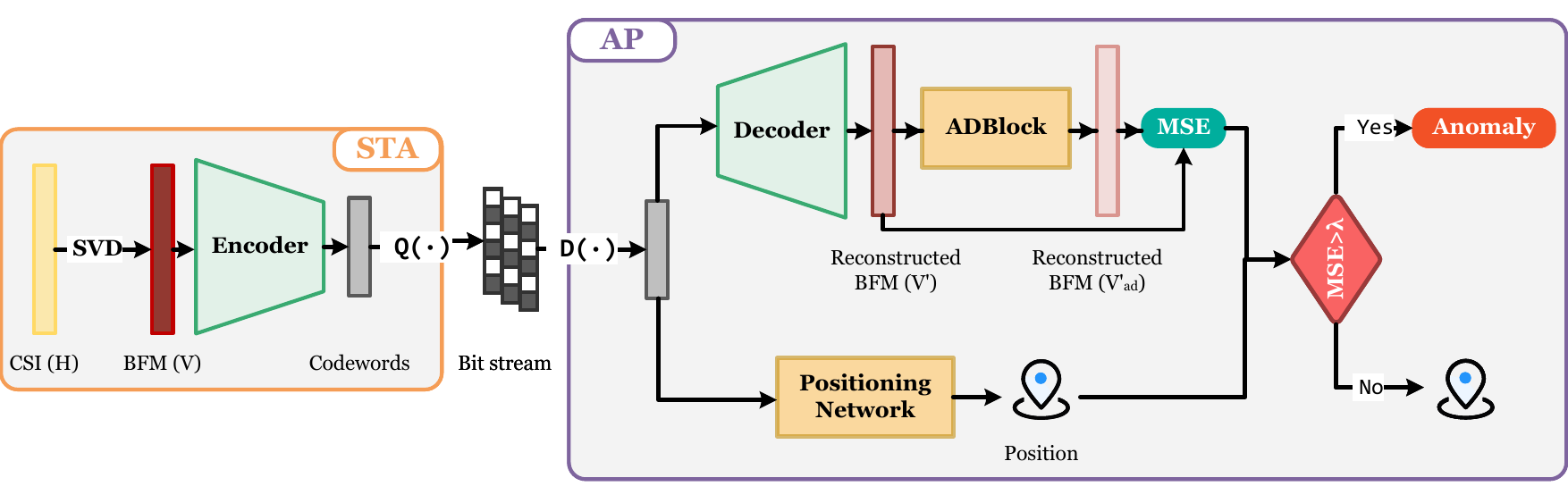}
    \caption{Overview of the FPNet framework for joint BFM feedback, positioning, and AD. At the STA, CSI $\mathbf{H}$ is decomposed into the BFM $\mathbf{V}$ via SVD, compressed into low-dimensional codewords, and then quantized into feedback bits. At the AP, the received bits are dequantized and processed by a decoder and a positioning network. The reconstructed BFM $\mathbf{V}'$ is also fed into ADBlock, which produces $\mathbf{V}'_{\mathrm{ad}}$. The MSE between $\mathbf{V}'$ and $\mathbf{V}'_{\mathrm{ad}}$ is compared to a threshold $\lambda$ to detect anomalies.}
    \label{fig:adblock}
    \vspace{-10pt}
\end{figure*}

Training a joint NN for multiple tasks can introduce task interference, especially when tasks share common feature representations. To mitigate this, we employ a two-step training procedure, as depicted in Algorithm~\ref{alg:fpnet_training}:

\begin{algorithm}[t]
\caption{Two-step training procedure of FPNet}
\label{alg:fpnet_training}
\begin{algorithmic}[1]
\State \textbf{Input:} Training data $\mathcal{D}$;
       initial parameters $(\theta_{\mathrm{enc}}^{(0)}, \theta_{\mathrm{dec}}^{(0)}, \theta_{\mathrm{pos}}^{(0)})$;
       hyperparameter $\alpha$;
       learning rates $\eta_1$, $\eta_2$.
\State \textbf{Output:} Trained FPNet parameters
       $(\theta_{\mathrm{enc}}^\star, \theta_{\mathrm{dec}}^\star, \theta_{\mathrm{pos}}^\star)$.
\State \textbf{Step 1: Positioning-oriented pre-training}
\State Initialize encoder and positioning network:
       $\theta_{\mathrm{enc}} \gets \theta_{\mathrm{enc}}^{(0)}$, 
       $\theta_{\mathrm{pos}} \gets \theta_{\mathrm{pos}}^{(0)}$.
\State Train encoder and positioning network by minimizing the positioning loss 
       $L_{\mathrm{pos}}$ in~\eqref{eq:position} using learning rate $\eta_1$, 
       until the neural network converges.
\State Save parameters after Step~1:
       $\theta_{\mathrm{enc}}^{\mathrm{saved}} \gets \theta_{\mathrm{enc}}$, 
       $\theta_{\mathrm{pos}}^{\mathrm{saved}} \gets \theta_{\mathrm{pos}}$.
\State \textbf{Step 2: End-to-end fine-tuning with combined loss}
\State Initialize full FPNet:
       $\theta_{\mathrm{enc}} \gets \theta_{\mathrm{enc}}^{\mathrm{saved}}$, \;
       $\theta_{\mathrm{pos}} \gets \theta_{\mathrm{pos}}^{\mathrm{saved}}$, \;
       $\theta_{\mathrm{dec}} \gets \theta_{\mathrm{dec}}^{(0)}$.
\State Train FPNet end-to-end by minimizing the combined loss
       \[
          L = L_{\mathrm{pos}} + \alpha L_{\mathrm{bfm}},
       \]
       using the reduced learning rate $\eta_2$ until convergence.
\State \Return $(\theta_{\mathrm{enc}}^\star, \theta_{\mathrm{dec}}^\star, \theta_{\mathrm{pos}}^\star)$.
\end{algorithmic}
\end{algorithm}

\begin{enumerate}
    \item \textbf{Step 1:} Train the encoder and positioning network jointly by minimizing the positioning loss in \eqref{eq:position}. Once the positioning performance is satisfactory, save the network parameters.

    \item \textbf{Step 2:}  Initialize the full model with the saved parameters and train end-to-end using the combined loss  
    \begin{equation}
        L = L_\mathrm{pos} + \alpha L_\mathrm{bfm}, \label{eq:both}
    \end{equation}
    where $\alpha$ balances positioning accuracy and BFM reconstruction. A reduced learning rate is used in this phase to fine-tune the parameters and ensure stable convergence.

\end{enumerate}

This stepwise approach simplifies training and ensures the encoder captures features critical for positioning. The composite loss allows the model to achieve both efficient BFM feedback compression and high positioning accuracy simultaneously.

\begin{algorithm}[t]
\caption{Training and deployment procedure of ADBlock}
\label{alg:adblock}
\begin{algorithmic}[1]
\State \textbf{Input:} 
       Training data $\mathcal{D}$ collected from the entire environment;
       normal training set $\mathcal{D}_\mathrm{norm}$ from known spatial locations;
       initial ADBlock parameters $\theta_{\mathrm{ad}}^{(0)}$;
       anomaly threshold $\tau$.
\State \textbf{Output:}
       Trained ADBlock parameters $\theta_{\mathrm{ad}}^\star$ and anomaly detection results on test data.
\State \textbf{Step 1: FPNet training}
\State Train FPNet using training data $\mathcal{D}$ collected from the entire environment, encompassing both normal and potentially anomalous samples, as described in Algorithm~\ref{alg:fpnet_training}, and obtain
       \[
       (\theta_{\mathrm{enc}}^\star, \theta_{\mathrm{dec}}^\star, \theta_{\mathrm{pos}}^\star).
       \]
\State \textbf{Step 2: ADBlock training}
\State Construct the ADBlock autoencoder at the AP and initialize its parameters:
       $\theta_{\mathrm{ad}} \gets \theta_{\mathrm{ad}}^{(0)}$.
\State Use the trained FPNet to generate reconstructed BFM outputs from the normal training set $\mathcal{D}_\mathrm{norm}$; for each clean sample, obtain $\mathbf{V}'$ from the FPNet decoder.
\State Train the ADBlock autoencoder on these reconstructed BFMs by minimizing its reconstruction loss (e.g., mean squared error between the input $\mathbf{V}'$ and the autoencoder output $\mathbf{V}'_{\mathrm{ad}}$) until the neural network converges, so that it learns the manifold of normal BFM reconstruction.
\State Save the trained ADBlock parameters:
       $\theta_{\mathrm{ad}}^\star \gets \theta_{\mathrm{ad}}$.
\State \textbf{Step 3: Anomaly detection on test data}
\For{each test BFM sample}
    \State Use the trained FPNet decoder to obtain the reconstructed BFM output $\mathbf{V}'$.
    \State Feed $\mathbf{V}'$ into the trained ADBlock autoencoder to obtain the secondary reconstruction $\mathbf{V}'_{\mathrm{ad}}$.
    \State Compute the reconstruction error using the mean squared error
           \[
               \mathrm{MSE} = \bigl\|\mathbf{V}' - \mathbf{V}'_{\mathrm{ad}}\bigr\|_2^2.
           \]
    \If{$\mathrm{MSE} > \lambda$}
        \State Flag the sample as anomalous.
    \Else
        \State Treat the sample as normal.
    \EndIf
\EndFor
\end{algorithmic}
\end{algorithm}

\subsection{AD by ADBlock}

To further enhance FPNet’s robustness in practical deployments, we introduce an AD module called ADBlock. While FPNet ensures efficient BFM compression and accurate positioning within the predefined region, real-world environments often produce BFM samples outside the training distribution, which can compromise positioning accuracy and thus call for a dedicated AD mechanism. The key motivation for ADBlock is that, under our feedback architecture, a \emph{single} autoencoder cannot provide usable anomaly scores at the AP: classical autoencoder-based AD relies on comparing the reconstruction with the original input, whereas in our system the encoder resides at the STA and the AP never observes raw CSI/BFM, making such a comparison impossible without breaking feedback compression. By adding a \emph{second} autoencoder (ADBlock) on top of FPNet’s decoder, both fully deployed at the AP, we enable anomaly scores to be computed solely from signals available at the AP, thereby integrating autoencoder-based AD into the compressed-feedback workflow.

The training and deployment of ADBlock consist of three steps, illustrated in Fig.~\ref{fig:adblock}, while the detailed procedure is given in Algorithm~\ref{alg:adblock}:

\textbf{Step 1: FPNet training.} First, the FPNet components (encoder, decoder, and positioning network) are trained as described in Section~\ref{sec:architecture}, without involving AD. In this phase, the training data include CSI samples collected from the entire environment, encompassing both normal and potentially anomalous samples. The FPNet model is trained in a supervised fashion using only location labels, without distinguishing between normal and anomalous inputs. Upon completion, this yields a model capable of accurate BFM feedback reconstruction and indoor positioning.

\textbf{Step 2: ADBlock training.} Once FPNet is trained, we construct and deploy the ADBlock module at the AP. ADBlock employs an autoencoder structure and is trained exclusively on normal BFM data from known, predefined spatial locations. Specifically, we use FPNet to generate reconstructed BFM outputs from clean training samples; these outputs serve as the training data for the ADBlock autoencoder, enabling it to learn the manifold of normal BFM reconstruction.

\textbf{Step 3: AD on test data.}  During inference, test BFM samples (normal and potentially anomalous) are first processed through the trained FPNet decoder to produce reconstructed BFM outputs $\mathbf{V}'$. These are then input to the trained ADBlock autoencoder, yielding secondary reconstructions $\mathbf{V}'_{\mathrm{ad}}$. The anomaly decision is based on the reconstruction error measured by the mean squared error:
\begin{equation}
    \mathrm{MSE} = \bigl\|\mathbf{V}' - \mathbf{V}'_{\mathrm{ad}}\bigr\|_2^2.
\end{equation}
If the MSE exceeds a predefined threshold, the sample is flagged as anomalous.

\subsection{Detailed NN Architecture}

\subsubsection{Encoder}
\begin{figure} 
    \centering
    \setlength{\abovecaptionskip}{-2mm}
    \includegraphics[width=0.99\linewidth,trim=1 0.7 1 3, clip]
    {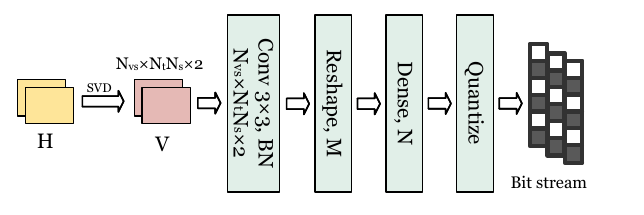}
    \caption{Encoder architecture. BFM is compressed and quantized to feedback bits.}
    \vspace{-12pt}
    \label{encoder}
\end{figure}

As illustrated in Fig.~\ref{encoder}, the real and imaginary parts of the BFM $\mathbf{V} \in \mathbb{C}^{N_\mathrm{vs} \times N_\mathrm{t}N_\mathrm{s}}$ are separated and treated as two input channels to the encoder. The input tensor is first passed through a convolutional layer with two $3\times 3$ kernels, followed by batch normalization (BN), to capture local spatial correlations while stabilizing the feature distribution for subsequent layers.
The resulting feature map is then flattened into a vector of length $M$, which is subsequently passed through a fully connected layer to produce a compact codeword of length $N$. Finally, this codeword is quantized using 5-bit uniform quantization and transmitted to the AP. 

\subsubsection{Decoder}
\begin{figure} 
    \centering
    \setlength{\abovecaptionskip}{2mm}
    \includegraphics[width=0.99\linewidth,trim=1 0.7 1 3, clip]
    {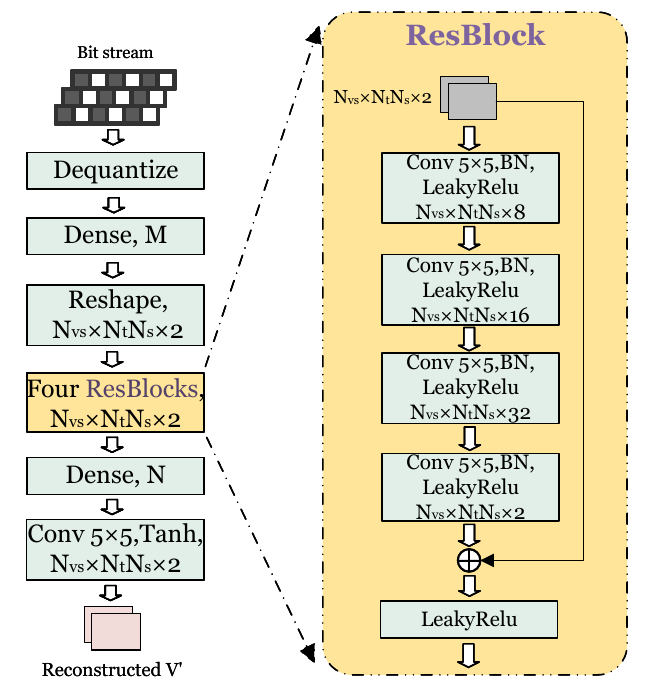}
    \caption{Decoder architecture. Four ResBlocks are used to enhance reconstruction performance.}
    \vspace{-8pt}
    \label{decoder}
\end{figure}

At the AP, the received quantized bitstream is dequantized to recover the compressed codeword, which is then input to the decoder, as depicted in Fig.~\ref{decoder}. The decoder begins with a fully connected layer followed by a reshaping operation to restore the original spatial dimensions. The reshaped tensor is then processed through a sequence of four ResBlocks, each designed to progressively refine reconstruction quality. Specifically, the ResBlocks consist of convolutional layers with 8, 16, 32, and 2 filters, respectively, each followed by a LeakyReLU activation. A final convolutional layer generates the fully reconstructed BFM.

\subsubsection{Positioning Network}
At the AP, the compressed BFM is also input into a positioning network consisting of a single fully connected layer. The final layer employs a softmax function to predict the STA’s position category: 
\begin{equation}   
    \text{softmax}(y_i) = \frac{e^{y_i}}{\sum_{j=1}^{C} e^{y_j}},
\end{equation}
where $C$ is the total number of classes, $y_i$ is the $i$-th input value to the softmax layer, and $\mathrm{softmax}(y_i)$ is the output probability for class $i$. All convolutional layers use zero-padding to maintain consistent spatial dimensions.

\subsubsection{ADBlock}
The ADBlock module is implemented as an autoencoder and deployed entirely at the AP. Unlike the main FPNet pipeline, ADBlock does not perform quantization or dequantization, since all computations occur locally at the AP. This reduces complexity and preserves information fidelity.

The architecture of ADBlock is summarized in Table~\ref{table:adblock}. 
The ADBlock encoder is \emph{identical} to the FPNet encoder: it processes the real and imaginary parts of the reconstructed BFM $\mathbf{V}' \in \mathbb{C}^{N_\mathrm{vs} \times N_\mathrm{t}N_\mathrm{s}}$ using a convolutional layer, flattening, and a fully connected layer to generate a low-dimensional latent representation. The decoder mirrors the FPNet decoder but with only two ResBlocks, each composed of convolutional layers followed by LeakyReLU activations, to ensure lightweight inference. A final convolutional layer produces the reconstructed output $\mathbf{V}'_{\mathrm{ad}}$.

ADBlock is trained using only normal BFM samples (i.e., reconstructed outputs $\mathbf{V}'$ from FPNet under known conditions) with the MSE as the reconstruction loss:
\begin{equation}
    L_{\mathrm{AD}} = \|\mathbf{V}' - \mathbf{V}'_{\mathrm{ad}}\|_2^2.
\end{equation}
During inference, $\mathbf{V}'$ is passed through the trained ADBlock, and the reconstruction error between $\mathbf{V}'$ and $\mathbf{V}'_{\mathrm{ad}}$ is computed. If the error exceeds a predefined threshold $\lambda$, the sample is flagged as anomalous, enabling robust detection of OOD BFM inputs.

\begin{table*}[t]
\caption{Architecture and Complexity Details of the ADBlock Module}
\label{table:adblock}
\renewcommand{\arraystretch}{1.2}
\begin{center}
\begin{tabular}{|l|l|l|l|}
\hline
\textbf{Layer} & \textbf{Output Shape} & \textbf{Parameter Number} & \textbf{FLOP Number} \\ \hline
\multicolumn{4}{|c|}{\textbf{Encoder (identical to FPNet)}} \\ \hline
Input: $\mathbf{V}'$ (real + imag) & $N_{\mathrm{vs}} \times N_{\mathrm{t}}N_{\mathrm{s}} \times 2$ & 0 & 0 \\ \hline
Conv ($3\times3$, 2 filters) & $N_{\mathrm{vs}} \times N_{\mathrm{t}}N_{\mathrm{s}} \times 2$ & $18$ & $18 \cdot N_{\mathrm{vs}} \cdot N_{\mathrm{t}}N_{\mathrm{s}}$ \\ \hline
Flatten             & $M \times 1$ & 0 & 0 \\ \hline
FC (latent)         & $N \times 1$ & $N(M+1)$ & $2NM$ \\ \hline

\multicolumn{4}{|c|}{\textbf{Decoder (2 ResBlocks)}} \\ \hline
FC + Reshape        & $N_{\mathrm{vs}} \times N_{\mathrm{t}}N_{\mathrm{s}} \times 2$ & $\sim$ & $\sim$ \\ \hline
ResBlock 1 & $N_{\mathrm{vs}} \times N_{\mathrm{t}}N_{\mathrm{s}} \times 2$ & $25 \cdot T_1$ & $2 \cdot 25 \cdot T_1 \cdot N_{\mathrm{vs}} \cdot N_{\mathrm{t}}N_{\mathrm{s}}$ \\ \hline
ResBlock 2 & $N_{\mathrm{vs}} \times N_{\mathrm{t}}N_{\mathrm{s}} \times 2$ & $25 \cdot T_1$ & $2 \cdot 25 \cdot T_1 \cdot N_{\mathrm{vs}} \cdot N_{\mathrm{t}}N_{\mathrm{s}}$ \\ \hline
Final Conv Layer & $N_{\mathrm{vs}} \times N_{\mathrm{t}}N_{\mathrm{s}} \times 2$ & $36$ & $36 \cdot N_{\mathrm{vs}} \cdot N_{\mathrm{t}}N_{\mathrm{s}}$ \\ \hline
\end{tabular}
\end{center}

\noindent\textit{Note:} $T_1 = 2\cdot8 + 8\cdot16 + 16\cdot32 + 32\cdot2 = 1648$ represents the total number of weights in one ResBlock.
\end{table*}

\section{Experiments and Results}

\subsection{Experiment Setup and Preparation}

\subsubsection{CSI Collection}
\begin{figure}[t]
  \centering
  \subfigure[STA]{
    \label{sta}
    \includegraphics[width=0.30\linewidth]{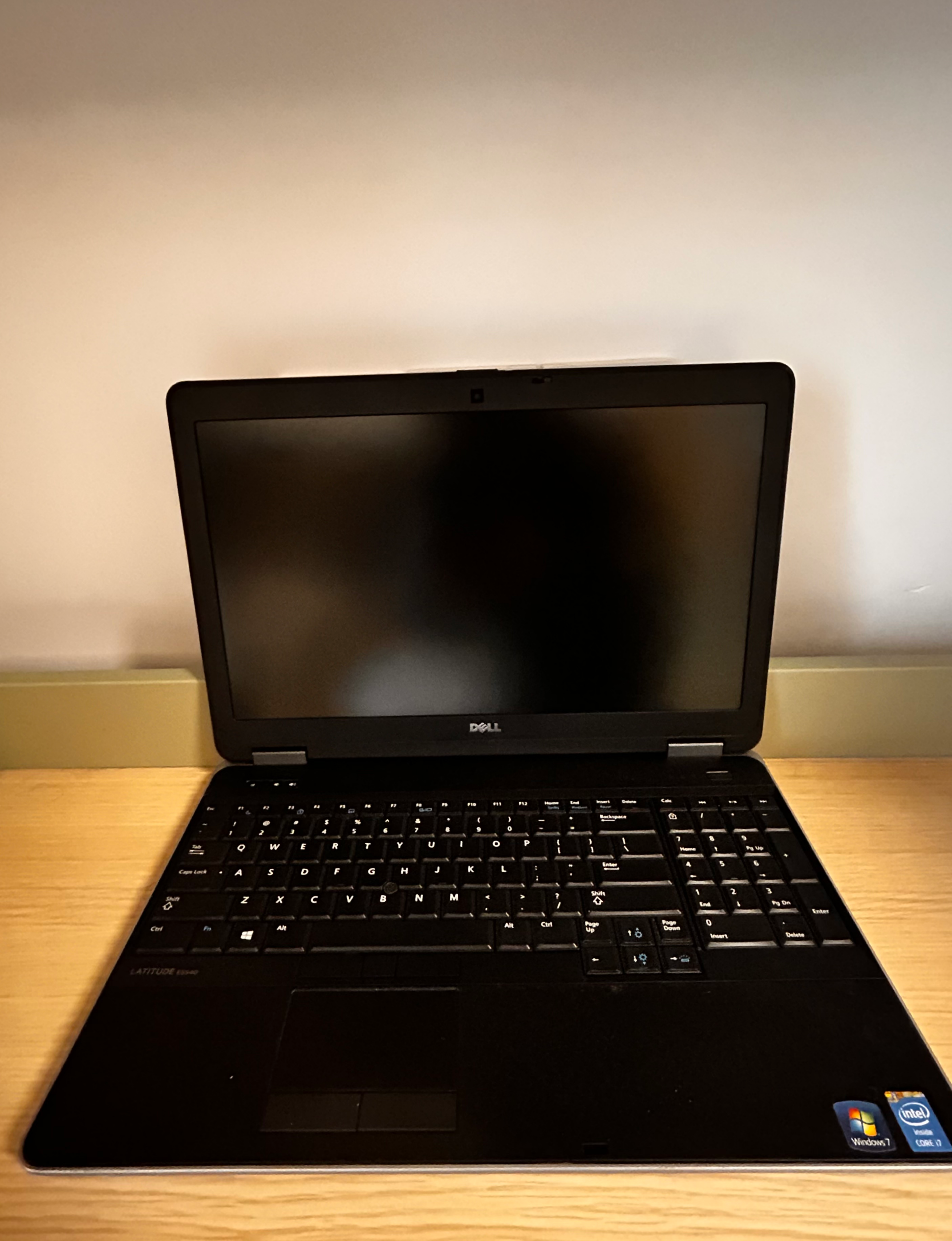}}
  \subfigure[STA]{
    \label{AP}
    \includegraphics[width=0.30\linewidth]{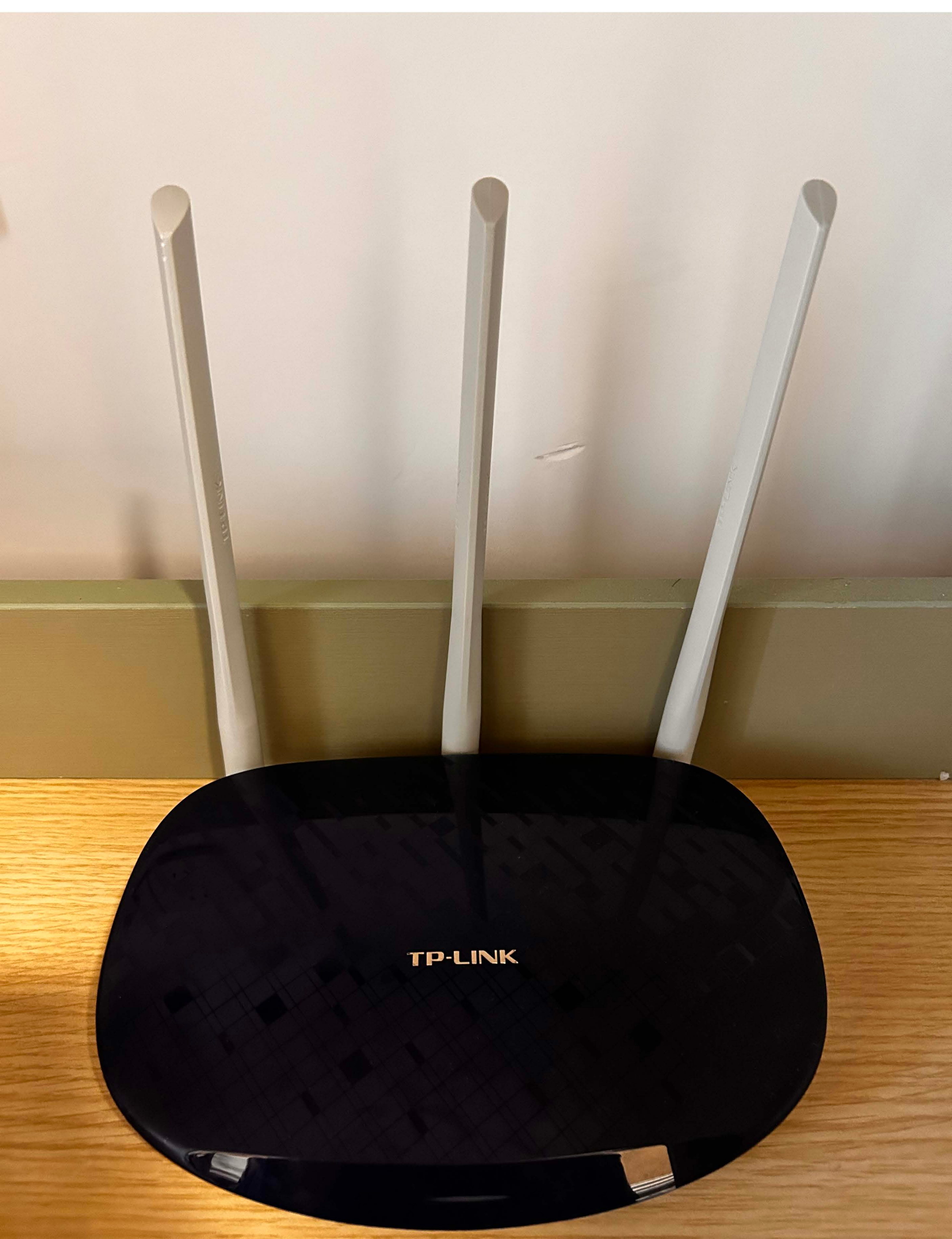}}
  \subfigure[Intel 5300 card]{
    \label{intel5300}
    \includegraphics[width=0.30\linewidth]{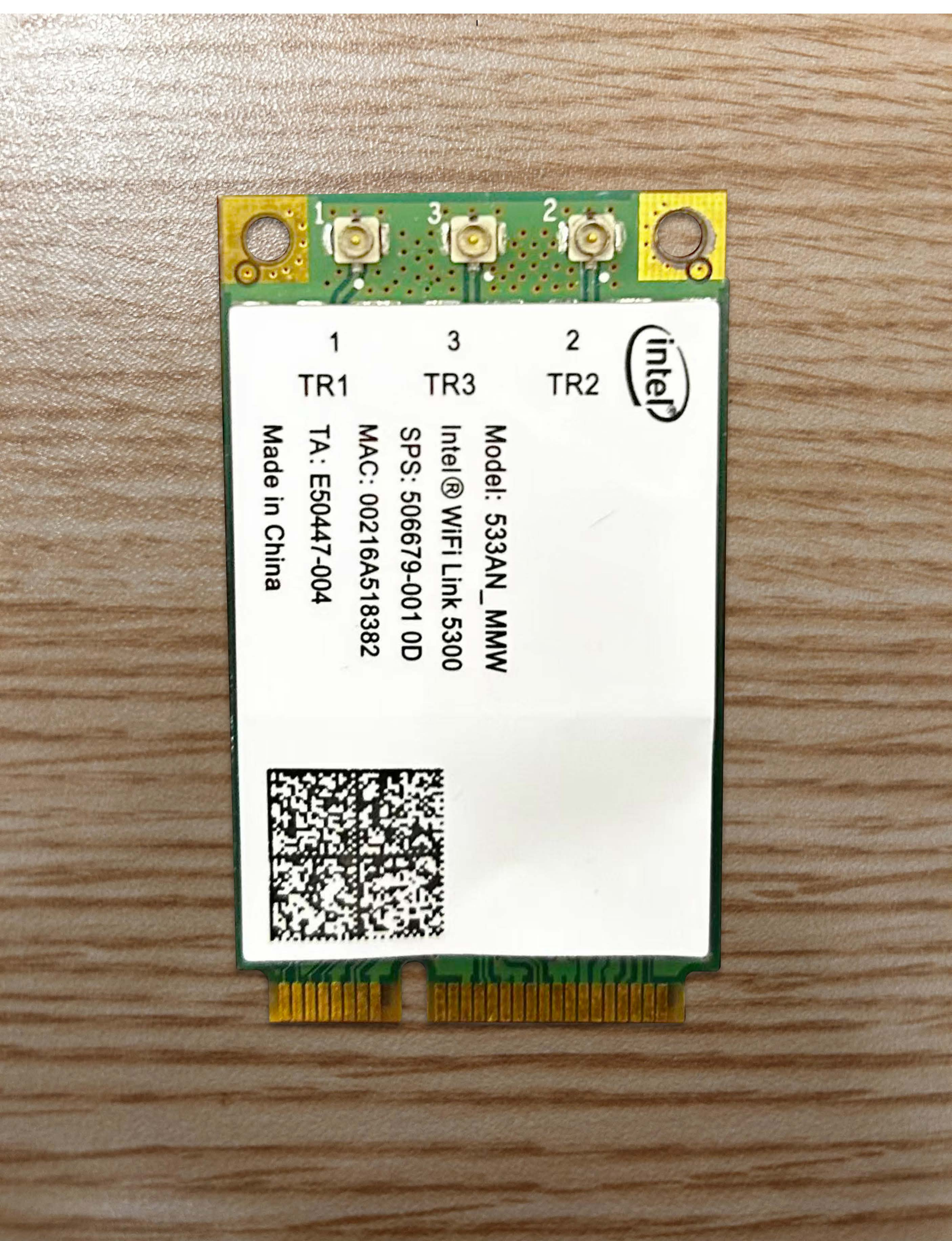}}
  \caption{Equipments used for the experiment.}
  \label{Devices}
  \vspace{-8pt}
\end{figure}

To evaluate the performance of the joint network FPNet, practical experiments were conducted using a Dell laptop, a TP-Link TL-WR886N router, an Intel 5300AGN wireless network card, and the CSI Tool platform. The router, equipped with two antennas, acted as the AP, while the laptop, fitted with the Intel 5300AGN card and three antennas, served as the STA. 
We select the 5300 platform primarily for fair comparison and reproducibility, as it has been widely used in prior CSI-based sensing/localization studies. All experiments are conducted in the 2.4\,GHz band with 20\,MHz channels; the tool reports CSI on 30 subcarrier groups, from which we retain \emph{28 valid subcarriers}. The hardware arrangement is shown in Fig.~\ref{Devices}.

While the Intel 5300 CSI Tool is a legacy 802.11n platform with limited subcarrier resolution, bandwidth, and spatial streams, it remains one of the most mature and widely used CSI capture systems in the literature~\cite{CSItool}, with well-validated hardware and driver behavior and publicly available datasets. In our design, FPNet is capture-agnostic. Replacing the Intel 5300-based front end with traces from modern platforms such as Nexmon or ath9k-based 802.11ac/ax sniffers or PicoScenes only requires adapting the input dimensions and normalization to the new CSI or BFM format~\cite{gringoli2019free,spachos2020esp32,liu2020picoscenes}, while keeping the core model architecture and training pipeline unchanged.

\begin{figure}[t]
  \centering
  \subfigure[Real scene]{
    \label{sta}
    \includegraphics[width=0.44\linewidth]{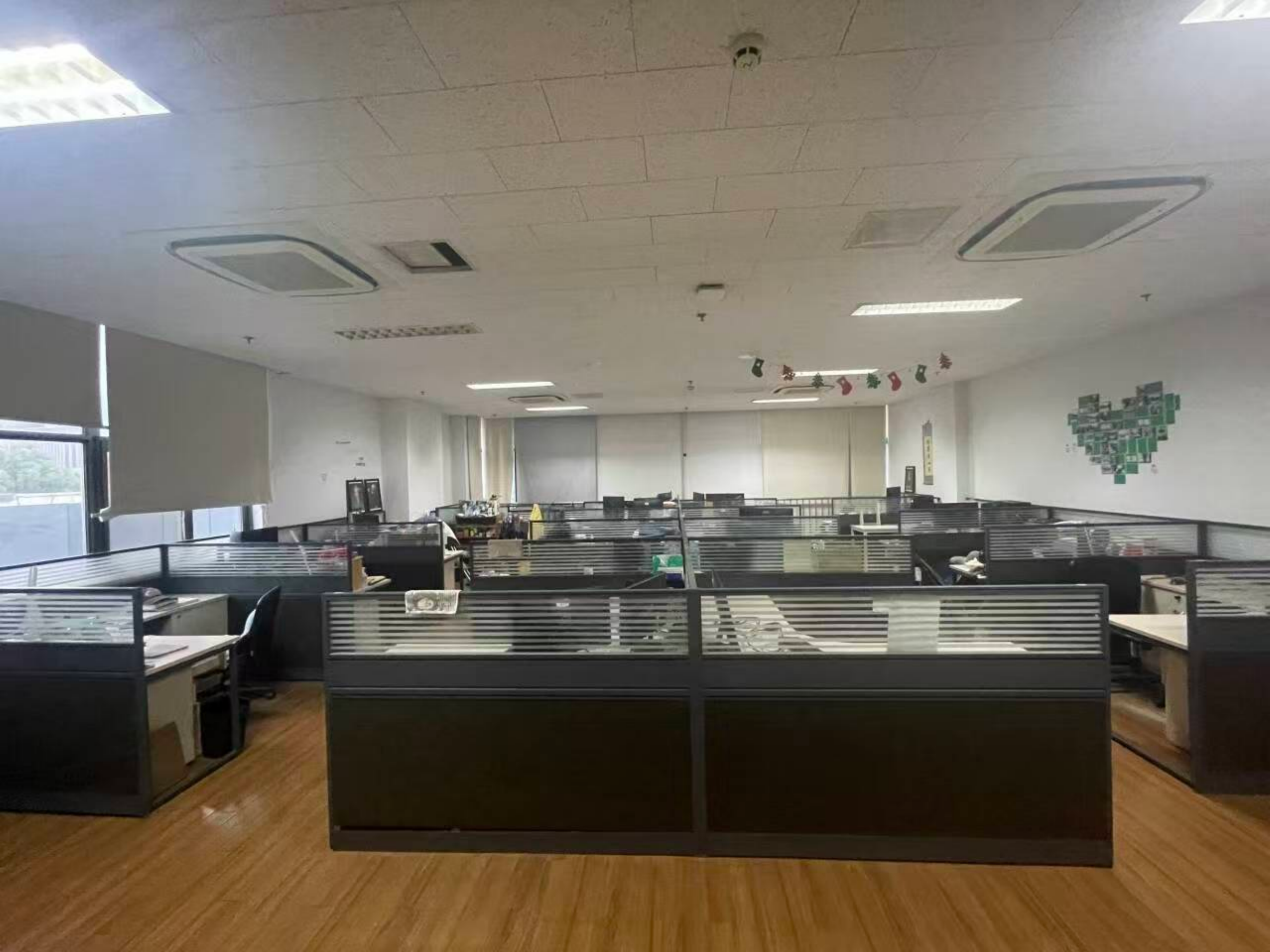}
    \vspace{9pt}
    }%
  \subfigure[Layout]{
    \label{sta}
    \includegraphics[width=0.50\linewidth ,trim=1 22 1 0.7, clip]{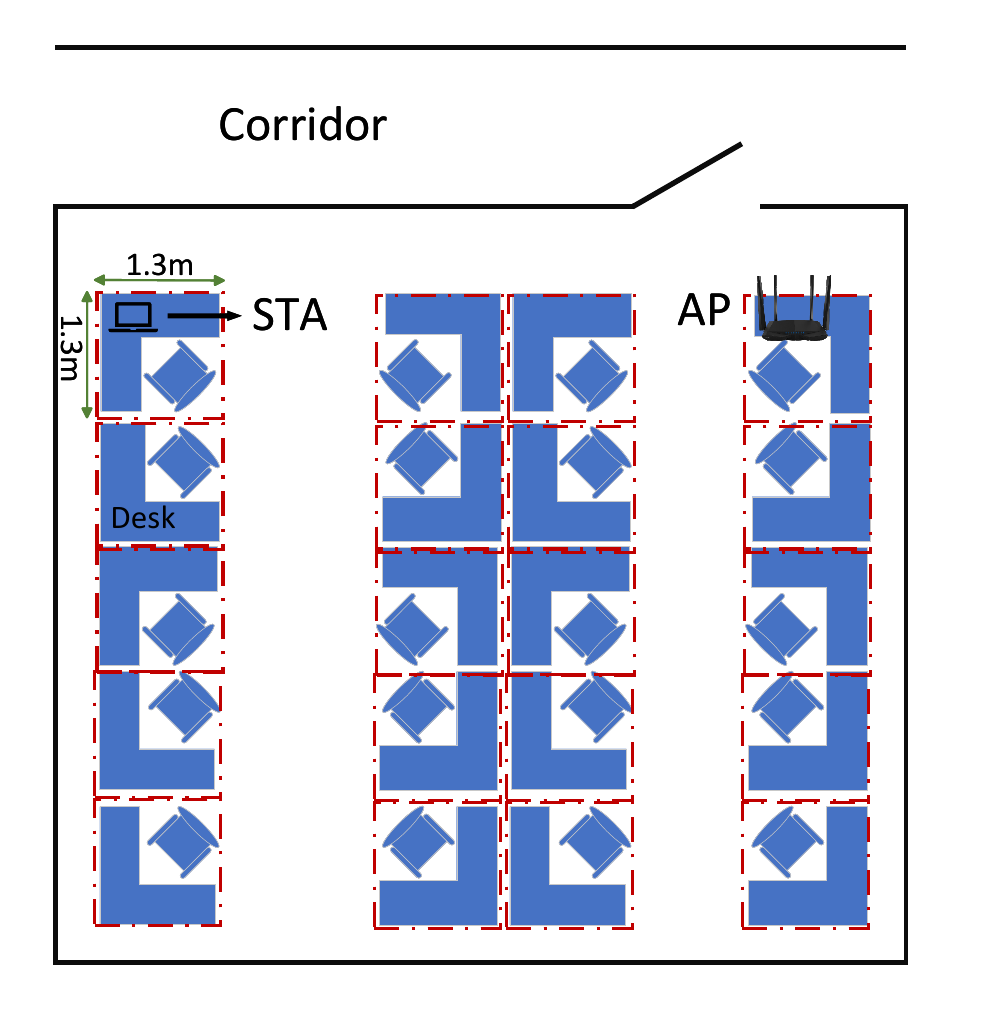}
    }
  \caption{Layout of the CSI data collection environment.}
  \vspace{-10pt}
  \label{fig:office_layout}
\end{figure}

The experiments took place in an indoor office environment measuring $8.2\times 9.3\, {\rm m}^2$, as depicted in Fig.~\ref{fig:office_layout}. The AP was positioned in the upper right corner of the space. CSI data were collected in each of the 20 delineated zones in the layout for one minute, capturing 5{,}000 packets per zone, with each zone covering approximately $1.3\times 1.3\,\mathrm{m}^2$.
The dataset was split into training, validation, and test sets in an 8:1:1 ratio.
In addition to the normal office CSI data, anomalous data were gathered in a corridor outside the office. This corridor dataset serves as an OOD set for training and evaluating the AD module (ADBlock), testing the system’s ability to distinguish anomalous samples from those in the office environment.

\subsubsection{Benchmark}
To comprehensively evaluate the performance of FPNet, we compare it against two categories of benchmarks: \textit{feedback-oriented methods} (including DL-based and protocol-based schemes) and \textit{traditional fingerprinting techniques}.

\textbf{Feedback-oriented Methods:}
To demonstrate that FPNet retains more position-relevant information through joint training, we compare it with three specific benchmarks in this category. These methods follow a two-stage sequential structure: first reconstructing the BFM, then performing positioning based on the reconstructed BFM.

\begin{itemize}
    \item \textbf{S-FPNet:} This benchmark is inspired by CFNet~\cite{he2025cfnet}, which compresses CSI and then applies the reconstructed CSI to downstream sensing tasks. Similarly, S-FPNet uses the same encoder and decoder architectures as FPNet to compress and reconstruct the BFM, but its positioning network is trained solely on the BFM, without joint optimization of feedback and positioning.
    
    \item \textbf{EFNet~\cite{qifan2024}:} This method employs an autoencoder with a channel attention mechanism to perform CSI feedback. In our implementation, we adapt the input and output dimensions to match the BFM format used in this work, and attach the same positioning network as in S-FPNet for subsequent position prediction.
    
    \item \textbf{Standard Protocols (T0/T1):} We also compare against the feedback schemes defined in the IEEE 802.11 MIMO standard. For clarity, we denote:
    \begin{itemize}
        \item \textbf{T0}: Type 0 angle-quantization feedback;
        \item \textbf{T1}: Type 1 angle-quantization feedback.
    \end{itemize}
    Both are implemented exactly as specified in~\cite{ieee80211ax2021}. No additional subcarrier grouping is applied, since the CSI Tool's acquisition already subsamples subcarriers, resulting in weak correlation between adjacent reported subcarriers.
\end{itemize}

For fair comparison, S-FPNet, EFNet, and the T0/T1 baselines all employ the same positioning network architecture (based on a modified LeNet) with input and output layers adapted to our data format.

\textbf{Traditional Fingerprinting Techniques:}
In addition to DL-based benchmarks, we incorporated traditional fingerprinting baselines to explicitly demonstrate the added value of the DL approach. We implemented k-Nearest Neighbors (KNN) and Support Vector Machine (SVM) using the same BFM dataset.
For KNN, we utilized distance-weighted voting with Euclidean distance.
For SVM, we evaluated both a linear kernel (LinearSVC) to test for linear separability and a Radial Basis Function (RBF) kernel to capture non-linear patterns.
Hyperparameters for both methods were optimized via grid search.

\subsubsection{NN Training}
All training and testing were conducted on an NVIDIA Tesla V100 GPU using the Adam optimizer. The batch size for all training processes was set to 64. For S-FPNet, training ran for 500 epochs with a learning rate of $5\times10^{-4}$. For FPNet, the first training phase also used 500 epochs and a learning rate of $5\times10^{-4}$, while the second phase used 300 epochs and a reduced learning rate of $10^{-4}$.

The ADBlock module was trained separately on normal BFM samples (i.e., outputs reconstructed by FPNet under known conditions). Training employed the Adam optimizer with a batch size of 64, a learning rate of $10^{-4}$, and 200 epochs to ensure the autoencoder effectively learned the normal BFM distribution. The reconstruction error, measured by  MSE, was minimized during training. The AD threshold $\lambda$ was empirically determined on the validation set to enable reliable OOD detection during inference.

\subsection{Performance Metrics}

\subsubsection{Evaluation Metrics for BFM Feedback}

For evaluating BFM reconstruction accuracy, we use the mean squared generalized cosine similarity (SGCS), defined as:
\begin{equation}   
    \rho^2={\frac1{N_\mathrm{vs}}\sum_{n=1}^{N_\mathrm{vs}}{ \left(\frac{|\widehat{\mathbf{v}}_n \mathbf{v}_n|}{\|\widehat{\mathbf{v}}_n\|_2\|\mathbf{v}_n\|_2} \right)^2}}
    .
\end{equation}  
where $\|\cdot\|_2$ is the Euclidean norm, $\widehat{\mathbf{v}}_n$ and $\mathbf{v}_n$ denote the $n$-th vector in reconstructed and original BFMs, respectively. While accurate BFM reconstruction is crucial for signal quality, excessive feedback overhead can negatively impact the transmission of effective data in communication systems. Thus, evaluating the actual throughput is equally important.

We simulate a MIMO transmission link, including precoding and modulation/demodulation, to calculate the error vector magnitude (EVM), which quantifies the similarity between transmitted and received signals: 
\begin{equation}   
\mathrm{EVM}=10 \cdot \log_{10}{ \frac{ \|\hat{\mathbf{x}}[k]-\mathbf{x}[k]\|_2^2}{\|\mathbf{x}[k]\|_2^2} }~\text{(dB)},
\end{equation}
where $\mathbf{x}[k]$ and $\hat{\mathbf{x}}[k]$ are the original and estimated signals. 

Gross throughput, representing the achievable data rate before accounting for BFM overhead, is given by 
\begin{equation}  
    R_\mathrm{gross}
    = \frac{N_{\mathrm{vs}}}{N_{\mathrm{fft}} + N_{\mathrm{cp}}}
      \times \mathrm{BW}
      \times \gamma(\mathrm{EVM}),
\end{equation}
where $N_{\mathrm{vs}}$, $N_{\mathrm{fft}}$, and $N_{\mathrm{cp}}$ are the numbers of valid subcarriers, fast fourier transform (FFT) points, and cyclic prefix length (28, 64, and 16, respectively), $\mathrm{BW}$ is the system bandwidth (40 MHz), and $\gamma(\mathrm{EVM})$ is the average number of bits per subcarrier determined by the EVM~\cite{IEEE802.11ac}.

Net throughput, reflecting the actual data rate after subtracting protocol overhead, is defined as 
\begin{equation} 
R_\mathrm{net} = R_\mathrm{gross}\times\frac{T}{T+T_{\mathrm{overhead}}},
\end{equation}
where $T$ is the data transmission time for a 300-byte packet and $T_{\mathrm{overhead}}$ is the fixed time for NDPA, NDP, ACK, and CBR frames. While $T_{\mathrm{overhead}}$ is constant, $T$ depends on the modulation and coding scheme.

\begin{table*}[t]
\centering
\scriptsize    
\renewcommand{\arraystretch}{1.3}
\caption{Comparison of SGCS, positioning accuracy, EVM, and throughput across methods with different feedback bit counts.}
\begin{tabular}{|l|ccccc|ccccc|c|c|c|}
\hline
Method & \multicolumn{5}{c|}{FPNet} & \multicolumn{5}{c|}{S-FPNet~\cite{he2025cfnet}} & EFNet~\cite{qifan2024} & T0~\cite{ieee80211ax2021} & T1~\cite{ieee80211ax2021} \\ \hline
Feedback bits & 60 & 70 & 80 & 90 & 100 & 60 & 70 & 80 & 90 & 100 & 100 & 672 & 896 \\ \hline
SGCS & 0.9172 & 0.9300 & 0.9422 & 0.9547 & 0.9585 & 0.9166 & 0.9246 & 0.9394 & 0.9523 & 0.9547 & 0.9441 & 0.9982 & \textbf{0.9996} \\
\rowcolor{Gray}
Positioning Accuracy & 96.23\% & 96.50\% & 96.61\% & 96.95\% & 97.52\% & 90.41\% & 91.47\% & 92.55\% & 94.38\% & 95.30\% & 92.81\% & 98.66\% & \textbf{98.70\%} \\
EVM (dB) & -14.15 & -14.63 & -15.01 & -15.61 & -16.28 & -14.14 & -14.48 & -14.64 & -15.36 & -15.83 & -15.09 & -20.61 & \textbf{-20.85} \\
$R_{\text{gross}}$ (Mb/s) & 21.5 & 21.5 & 21.5 & 21.5 & 28.0 & 21.5 & 21.5 & 21.5 & 21.5 & 21.5 & 21.5 & \textbf{42.0} & 42.0 \\
\rowcolor{Gray}
$R_{\text{net}}$ (Mb/s) & 9.64 & 9.59 & 9.54 & 9.48 & \textbf{10.35} & 9.64 & 9.59 & 9.54 & 9.48 & 9.43 & 9.43 & 8.42 & 7.57 \\ \hline
\end{tabular}
\label{results_table}
\end{table*}

\begin{figure}[t]
    \centering
    \setlength{\abovecaptionskip}{-2mm}
    \includegraphics[width=1.00\linewidth,trim=5 210 6 250, clip]   %
    {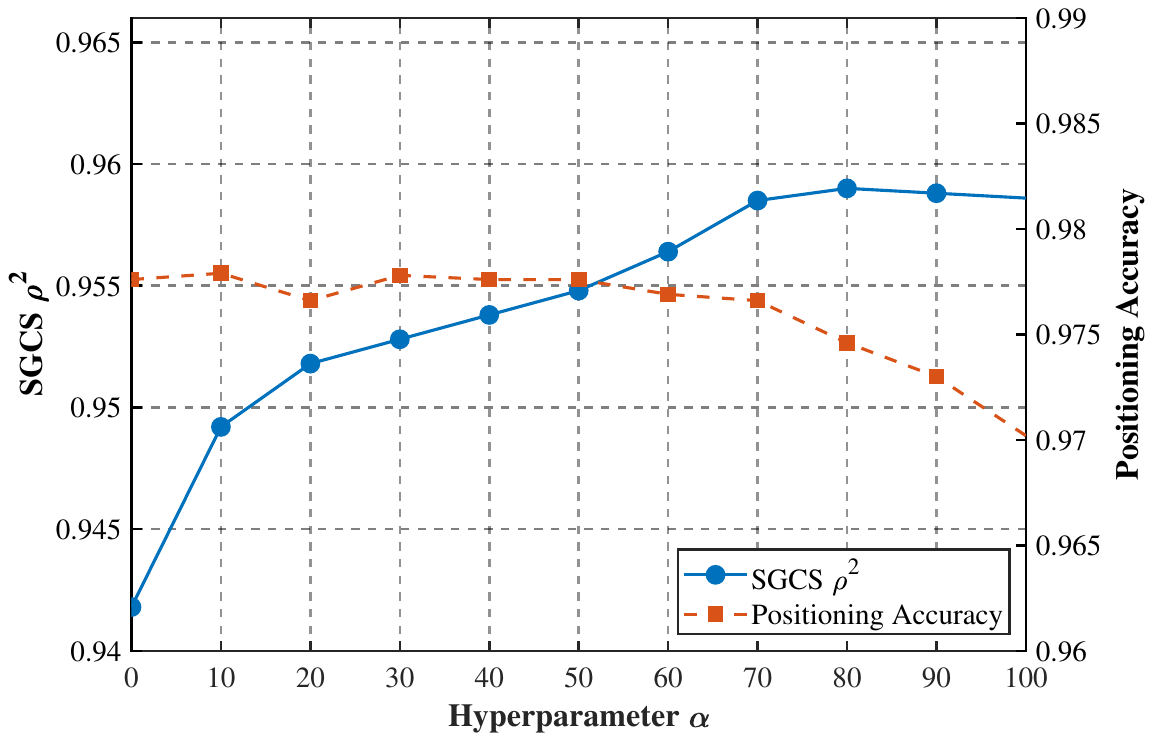}
    \caption{Trend of SGCS and positioning accuracy with varying hyperparameter $\alpha$.}
    \label{hyper}
\end{figure}

\subsubsection{Evaluation Metric for Positioning}
We simplify the positioning problem to a classification task, reflecting practical indoor positioning requirements. The environment is discretized into predefined location categories, and the objective is to classify the user's position into one of these categories.
Therefore, we use \emph{positioning accuracy}, the percentage of correctly classified samples, as the performance metric. This measure indicates how effectively the system predicts the correct location among the predefined categories; higher accuracy signifies a more accurate mapping from BFM data to location categories.

\subsubsection{Evaluation Metric for AD}

To evaluate the AD module, we use the following metrics:
\begin{itemize}
    \item \textbf{True Positives (TP):} The number of anomalous samples correctly flagged as anomalies.
    \item \textbf{False Positives (FP):} The number of normal samples incorrectly flagged as anomalies.
    \item \textbf{True Negatives (TN):} The number of normal samples correctly identified as normal.
    \item \textbf{False Negatives (FN):} The number of anomalous samples incorrectly identified as normal.
\end{itemize} 

From these, we compute:
\begin{itemize}
  \item \textbf{True Positive Rate (TPR)} (Recall):
  \[
    \mathrm{TPR} = \frac{\mathrm{TP}}{\mathrm{TP} + \mathrm{FN}}.
  \]
  \item \textbf{False Positive Rate (FPR)}:
  \[
    \mathrm{FPR} = \frac{\mathrm{FP}}{\mathrm{FP} + \mathrm{TN}}.
  \]
  \item \textbf{Precision}:
  \[
    \mathrm{Precision} = \frac{\mathrm{TP}}{\mathrm{TP} + \mathrm{FP}}.
  \]
  \item \textbf{F1-score}, the harmonic mean of Precision and Recall:
  \[
    \mathrm{F1\!-\!score} = 2 \times \frac{\mathrm{Precision} \times \mathrm{TPR}}
                                 {\mathrm{Precision} + \mathrm{TPR}}.
  \]
\end{itemize}
These metrics together provide a comprehensive assessment of ADBlock’s ability to distinguish anomalous from normal samples.

\subsection{Result Analysis}

\subsubsection{FPNet Result Analysis}
We first conducted experiments to determine the optimal value of the hyperparameter $\alpha$. As shown in Fig.~\ref{hyper}, increasing $\alpha$ causes the network to prioritize the BFM reconstruction task, leading to a continuous rise in SGCS. However, this focus comes at the expense of positioning performance, which declines as $\alpha$ grows. Eventually, SGCS plateaus, indicating an imbalance between the two tasks. Therefore, for subsequent experiments, we fixed $\alpha$ at 70. 
It is worth noting that $\alpha$ acts as a critical scaling factor to balance the gradient magnitudes of the two tasks, which differ in numerical scales. Our experiments on environment transfer (detailed later in the dynamic scenario analysis) reveal that $\alpha$ exhibits strong robustness: effective adaptation to the dynamic environment was achieved via fine-tuning without re-searching for $\alpha$. This suggests that re-tuning $\alpha$ is not required for moderate environmental changes, as the relative scale of reconstruction and positioning losses remains comparable, and grid search is likely only necessary under drastic distributional shifts.

Next, we compared DL-based feedback to protocol-based methods. Table~\ref{results_table} shows that as the number of feedback bits increases, S-FPNet’s BFM reconstruction accuracy, as measured by SGCS, also improves. EFNet attains reconstruction and throughput performance comparable to S-FPNet under the same feedback budget, though its performance is slightly weaker overall. The T0/T1 methods, which use a higher number of feedback bits, achieve superior reconstruction accuracy compared to DL-based approaches. This yields better EVM performance in simulations and a gross throughput nearly double that of S-FPNet. However, the significant feedback overhead of T0/T1 consumes a substantial portion of transmission time, reducing net throughput compared to S-FPNet and highlighting the efficacy of DL in feedback tasks.
 
We then evaluated the benefits of FPNet’s end-to-end multi-task training. Under the same feedback-bit conditions, FPNet outperforms S-FPNet in BFM reconstruction, although its accuracy remains below that of protocol-based methods. With 100 feedback bits, FPNet achieves higher net throughput, 9.73\% and 22.92\% improvements over S-FPNet and T0, respectively, demonstrating the potential of DL-based feedback and the advantages of joint training.

In positioning performance, FPNet shows marked improvements over S-FPNet. While S-FPNet’s positioning accuracy lags behind T0/T1 due to lower BFM reconstruction accuracy, FPNet surpasses S-FPNet. At 100 feedback bits, FPNet achieves 97.52\% positioning accuracy, confirming that joint multi-task training effectively compresses the BFM while preserving critical position-related information.

\begin{figure*}[t]
 \centering
\subfigure[SGCS versus epoch] 
{
\label{fig:SGCS}
\includegraphics[width=0.32\linewidth,trim=30 210 50 220, clip ]{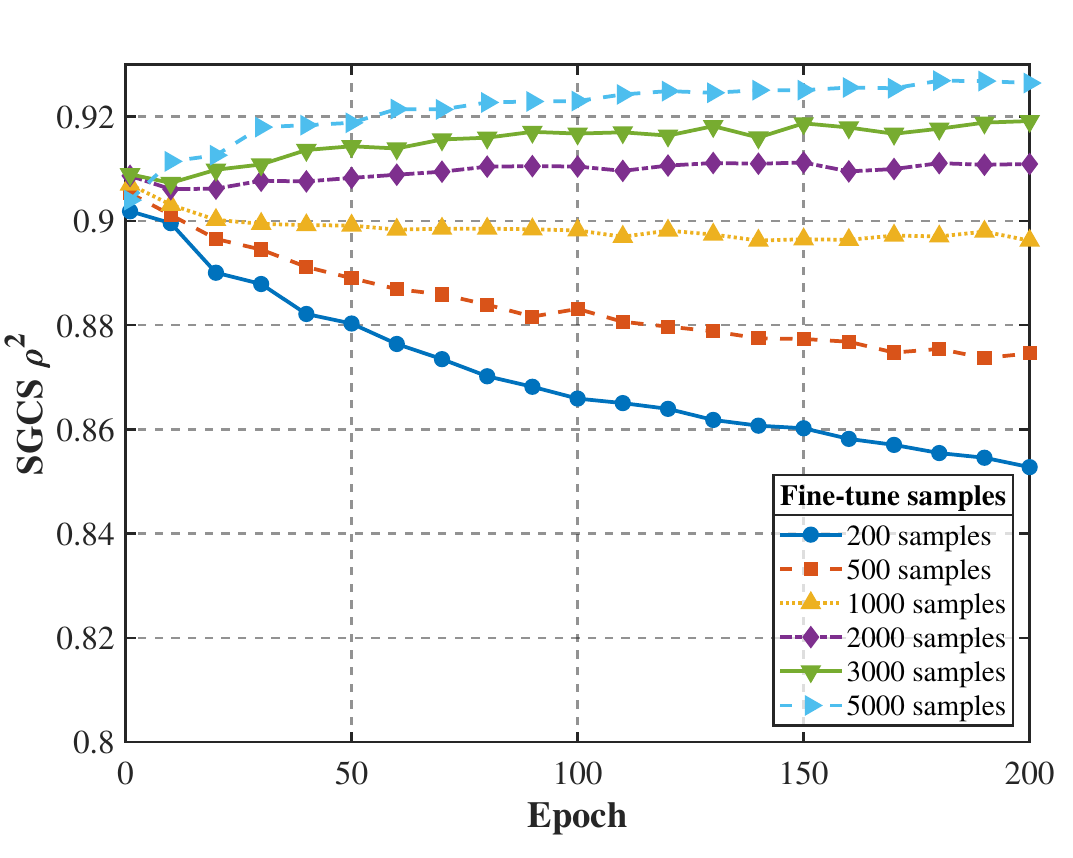}}
\subfigure[Positioning accuracy versus epoch] {
\label{fig:Positioning accuracy}
\includegraphics[width=0.32\linewidth,trim=30 210 50 220, clip ]{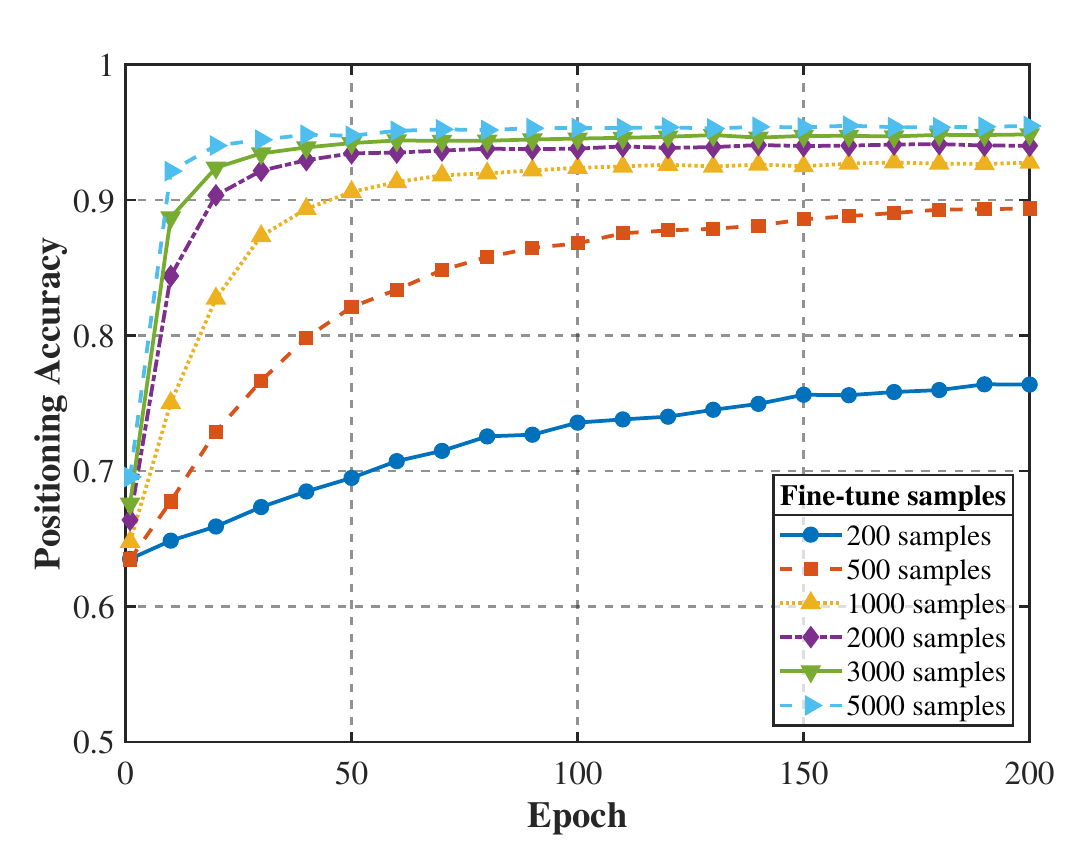}}
\subfigure[Training loss versus epoch] {\label{fig:training_loss}
\includegraphics[width=0.32\linewidth,trim=30 210 50 220, clip ]{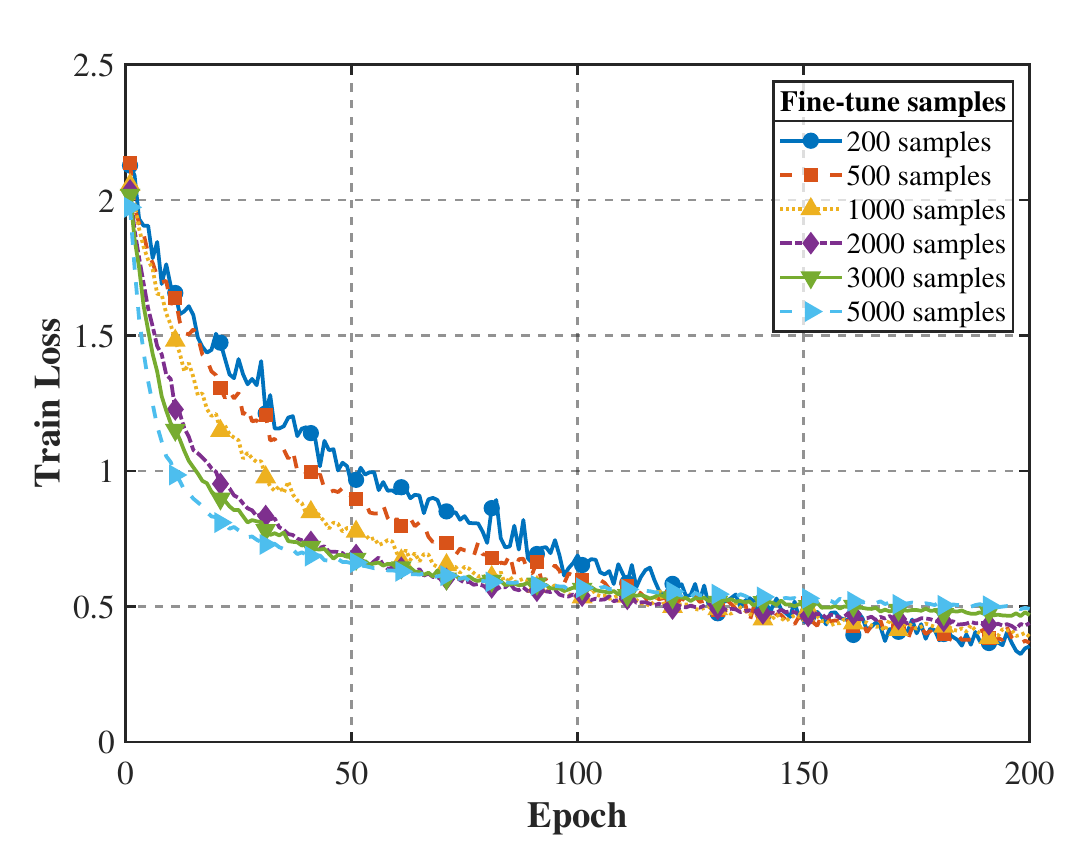}}
\caption{Fine-tuning performance under different fine-tune sample sizes. In (a) and (b), using more fine-tuning samples leads to steadily improving test performance as the number of epochs increases, whereas insufficient fine-tuning data results in slow gains or even degradation on the test set.
In (c), when the number of fine-tuning samples is too small, the loss converges slowly and may eventually lead to overfitting, whereas sufficient fine-tuning data are required for stable convergence and for the target performance to be reached.}
\label{fig:ENV_change}
\vspace{-10pt}
\end{figure*}

A key advantage of the proposed FPNet framework is that positioning is performed directly from the \emph{compressed BFM}, which is already required by the IEEE~802.11 feedback procedure. As a result, FPNet remains fully compatible with the Wi-Fi standard and introduces \emph{no additional feedback overhead}. Although the STA could, in principle, estimate its own position locally and send only the location label to the AP, this design has two practical drawbacks:

\begin{itemize}

  \item \textbf{Extra feedback overhead:} In our setup with 20 location classes, at least 5 additional bits are needed per packet. For example, when the BFM feedback budget is 100~bits, appending a 5-bit location label reduces the net throughput from 10.35~Mb/s to 10.32~Mb/s.
  \item \textbf{Increased device-side complexity:} If deep learning-based positioning is executed on the STA, the resulting computation and latency become non-negligible for resource-constrained user devices, as further discussed in \emph{Complexity and Running Time on FPGA}.
\end{itemize}

\subsubsection{Impact of Positioning Methods}
To validate the effectiveness of the proposed DL-based approach, we compared the positioning module of S-FPNet against the optimized KNN and SVM baselines described in Section IV-A.
Experimental results show that the Linear SVM failed to capture the data distribution, achieving an accuracy of approximately $\approx 25\%$, which indicates the strong nonlinearity of the BFM fingerprints.
Introducing the RBF kernel substantially improved the SVM accuracy to 97.97\%. The proposed deep learning based positioning network achieved the highest accuracy of 98.72\%, outperforming both KNN at 96.30\% and the RBF-SVM. These results confirm that the end-to-end feature learning capability of the DL model provides a clear performance advantage over traditional methods, particularly in extracting discriminative features from high-dimensional BFM data.

\subsubsection{Impact of the Number of Zones on FPNet Performance}

\begin{table}[t]
  \centering
  \caption{Effect of the number of zones on FPNet performance.}
  \label{tab:acc_vs_classes}
  \begin{tabular}{@{}ccc@{}}
    \toprule
    Number of Classes & Positioning Accuracy & SGCS \\
    \midrule
    5  & 98.11\% & 0.9580 \\
    10 & 97.95\% & 0.9573 \\
    20 & 97.52\% & 0.9585 \\
    40 & 94.19\% & 0.9561 \\
    \bottomrule
  \end{tabular}
\end{table}

\begin{table}[t]
  \centering
  \caption{FPNet performance in static and dynamic environments under different feedback bit counts.}
  \label{tab:static_dynamic_bits}
  \begin{tabular}{@{}ccccc@{}}
    \toprule
    Scenario & Metric & 80 bits & 90 bits & 100 bits \\
    \midrule
    \multirow{2}{*}{Static} 
      & SGCS                      & 0.9422 & 0.9547 & 0.9585 \\
      & Positioning Accuracy  & 96.61\%  & 96.85\%  & 97.52\%  \\
    \midrule
    \multirow{2}{*}{\shortstack{Dynamic\\(moving people)}} 
      & SGCS                      & 0.9184 & 0.9255 & 0.9331 \\
      & Positioning Accuracy  & 94.28\%  & 94.79\%  & 95.41\%  \\
    \bottomrule
  \end{tabular}
\end{table}

To further investigate the impact of the \textit{number of zones} on FPNet performance, we followed the same training pipeline while partitioning the office into 5, 10, 20, and 40 fine-grained subregions. The configuration with 20 subregions is identical to the setup described in the data collection, the coarser partitions (5 and 10 zones) are obtained by merging adjacent subregions, and the 40-zone configuration is constructed by further splitting each of the 20 original subregions into two.
From Table~\ref{tab:acc_vs_classes}, as the number of zones increases from 5 to 40, the BFM reconstruction metric (SGCS) changes only mildly, whereas positioning accuracy drops more noticeably (from 98.11\% to 94.19\%). This trend is expected: SGCS is label-agnostic and primarily reflects the fidelity of BFM reconstruction over the feedback distribution; as such, it is largely insensitive to how the space is partitioned. By comparison, finer partitions shrink inter-class margins while preserving strong CSI/BFM correlation among adjacent zones, which increases the likelihood of boundary ambiguities and thus misclassification. Although a finer spatial partition (to reduce distance error) is desirable for positioning accuracy, it not only increases training difficulty but also imposes substantial challenges on data collection
\cite{Wang2016DeepFi}.

\subsubsection{Impact of Environment Variation}

Since inferring positions from spatial signal patterns is sensitive to environmental dynamics, it is necessary to examine how changes in the propagation environment affect FPNet's performance. In practice, moving desks, chairs, and people walking around can substantially alter the wireless channel characteristics.
In addition to the data collection described in Section~IV-A (with surrounding objects kept static), we re-collected data at the same locations under a markedly changed \emph{dynamic environment} by allowing more people to walk and work inside the office.

As shown in Table~\ref{tab:static_dynamic_bits}, increasing the feedback bit count from 80 to 100 consistently improves both SGCS and positioning accuracy in both scenarios, confirming that FPNet benefits from a higher feedback budget. However, the dynamic environment with moving people exhibits a moderate degradation compared to the static case, with slightly lower SGCS and positioning accuracy across all bit counts.  This indicates that FPNet remains reasonably robust under channel perturbations induced by human motion, while still being sensitive to environment dynamics.

Applying the FPNet trained in the static scenario directly to the dynamic scenario reveals a clear divergence between the two tasks: while SGCS decreases only slightly from 0.9585 to 0.9055, positioning accuracy drops from 97.52\% to 64.59\% (noting that a randomly initialized network reaches only about 5\%). This indicates that FPNet preserves part of its positioning capability after cross-environment deployment, yet suffers some performance deterioration, whereas the BFM reconstruction task remains comparatively robust.
The limited degradation in SGCS can be explained by the intrinsic structure of the BFM. As established in the 802.11ac MU-MIMO sounding procedure and further clarified in BFMSense \cite{yi2024bfmsense}, the BFM is constructed from the dominant right singular vectors of the CSI matrix obtained through SVD, as expressed in \eqref{eq:SVD}. These singular vectors capture the principal spatial eigen-directions of the channel, which are governed mainly by the AP-STA antenna geometry and the large-scale angular structure of propagation. Because such geometric components evolve smoothly and remain nearly unchanged unless the transceiver configuration itself is modified, the BFM and its reconstructed representation naturally exhibit strong robustness to moderate environmental perturbations.

In contrast, even subtle environmental changes can be detrimental to position inference. The key reason lies in the fact that position classification relies heavily on fine-grained multipath patterns that serve as environment-specific spatial fingerprints. While the dominant spatial subspace remains stable, these small-scale multipath components are extremely sensitive to human motion, furniture displacement, and minor variations in reflectors. Prior studies consistently report that such high-resolution fingerprints are easily distorted by small perturbations, which substantially reshape the underlying CSI manifold and render models trained in one environment ineffective in another \cite{gao2019crisloc, ma2019wifi}. This pronounced sensitivity to small-scale fading explains why the positioning branch of FPNet experiences a dramatic accuracy collapse following an environment change, despite the relatively stable performance of BFM reconstruction.

To mitigate the severe degradation in positioning accuracy observed after cross-environment deployment, we adopt a fine-tuning strategy that adapts FPNet to the dynamic scenario using a small number of environment-specific samples. Fig. \ref{fig:ENV_change} summarizes the fine-tuning behavior under different sample sizes. Fig. \ref{fig:SGCS} and Fig. \ref{fig:Positioning accuracy} present the performance on the  test set immediately after each fine-tuning epoch. After few training epoches, the performance begins to rise steadily as FPNet gradually aligns its latent space with the new environment. The extent of improvement, however, depends strongly on the number of fine-tuning samples: Models fine-tuned with fewer than 1{,}000 samples show only marginal gains in positioning accuracy, and even suffer degraded BFM reconstruction performance on the test set due to overfitting, whereas sample sizes of 2{,}000 or more enable rapid recovery and allow quick convergence. When 5{,}000 samples are used for fine-tuning, the model achieves effective adaptation, with SGCS recovering to 0.9232 and positioning accuracy reaching 95.28\% after 100 epochs in the new environment.

Fig. \ref{fig:training_loss} illustrates the evolution of the training loss throughout the fine-tuning process. When the sample size is small, the loss decreases slowly, indicating that the model cannot sufficiently reshape its representation using limited data. As the number of fine-tuning samples increases, the loss converges faster, confirming that richer environment-specific data provide a more complete view of the updated multipath structure and facilitate stable adaptation. Together, these results demonstrate that fine-tuning is an effective means to restore FPNet’s performance under environmental changes, provided that a sufficient amount of representative samples from the new environment is available.

\begin{table}[t]
\centering
\caption{Distribution of anomalous BFMs predicted as each in-range position (without ADBlock).}
\label{tab:ood_distribution}
\renewcommand{\arraystretch}{1.2}
\setlength{\tabcolsep}{4pt}
\begin{tabular}{|c|cccccccccc|}
\hline
Pos        & 0   & 1   & 2   & 3   & 4   & 5   & 6   & 7   & 8   & 9   \\ \hline
Ratio (\%) & 4.1 & 9.0 & 7.1 & 5.6 & 4.5 & 4.1 & 8.6 & 5.4 & 5.6 & 3.0 \\ \hline
Pos        & 10  & 11  & 12  & 13  & 14  & 15  & 16  & 17  & 18  & 19  \\ \hline
Ratio (\%) & 4.5 & 2.6 & 2.2 & 6.0 & 3.7 & 2.2 & 1.9 & 6.2 & 6.0 & 7.7 \\ \hline
\end{tabular}

\begin{tablenotes}
\footnotesize
\item \textbf{Note:} Ratio (\%) indicates the percentage of anomalous BFM samples (i.e., out-of-distribution data) that were misclassified as each in-range position without applying the ADBlock module.
\end{tablenotes}
\end{table}

\begin{figure}[t]
    \centering
    \setlength{\abovecaptionskip}{-2mm}
    \includegraphics[width=1.00\linewidth,trim=5 210 6 250, clip]   
    {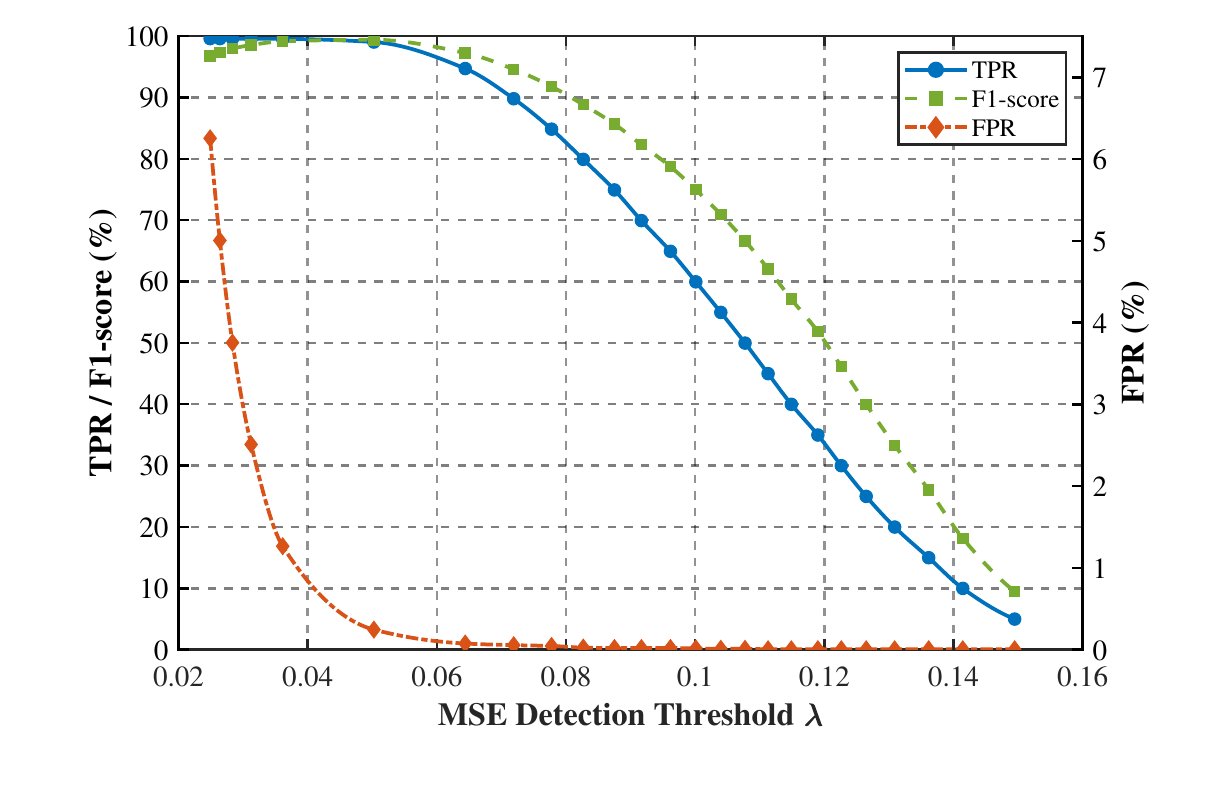}
    \caption{Detection performance curves under varying MSE threshold $\lambda$, illustrating the trade-off between sensitivity and false alarms. As $\lambda$ decreases, TPR increases but FPR also rises. The F1-score peaks at the optimal balance between recall and precision.}
    \label{fig:detection_metrics}
    \vspace{-10pt}
\end{figure}

\begin{figure}[t]
    \centering
    \setlength{\abovecaptionskip}{-2mm}
    \includegraphics[width=1.00\linewidth,trim=5 210 6 250, clip]   
    {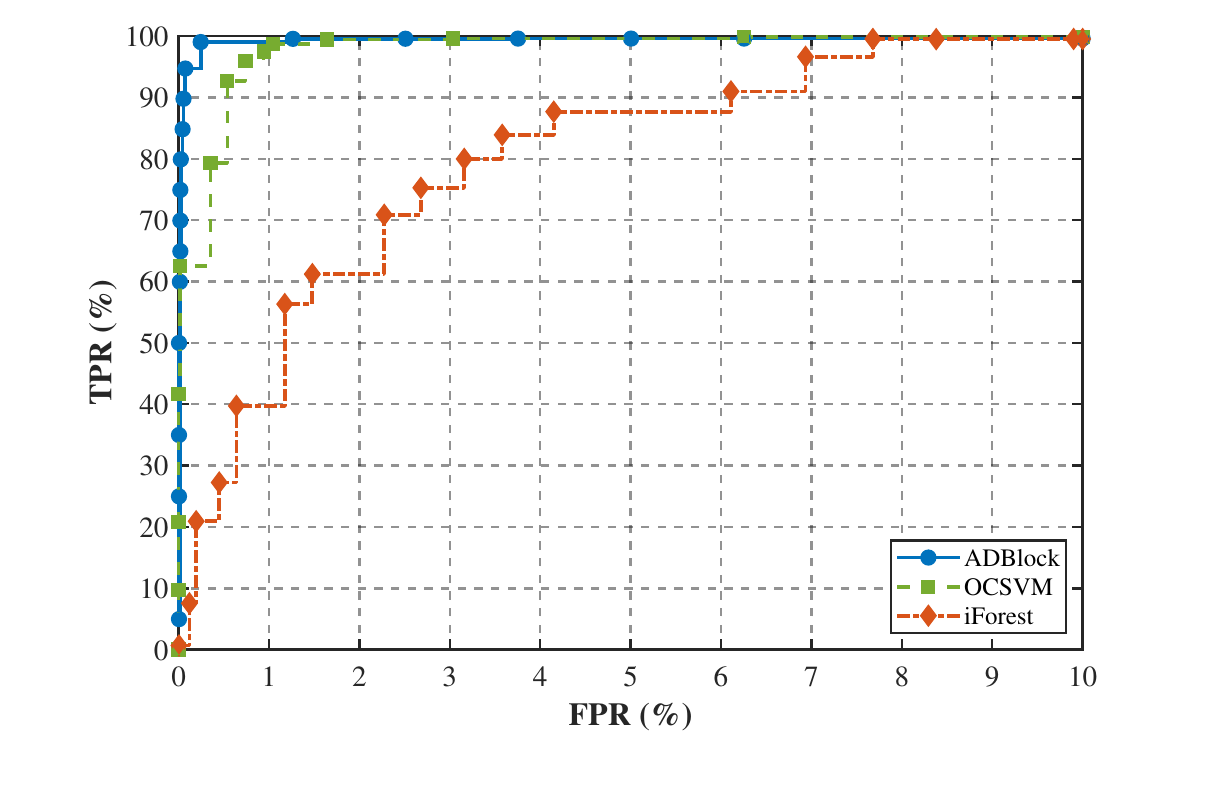}
    \caption{Threshold-swept ROC curves (TPR vs. FPR) in the low-FPR regime for anomaly detection on raw BFM fingerprints, comparing ADBlock, OCSVM, and iForest. All methods are trained using only normal samples and evaluated on the same mixed test set containing both normal samples and anomalies.}
    \label{fig:AD_compare}
    \vspace{-10pt}
\end{figure}

\subsubsection{ADBlock Performance Analysis}
We demonstrate the necessity of explicit anomaly detection by first evaluating the trained FPNet on a mixed test set comprising both normal and anomalous BFM samples, without applying ADBlock. Consequently, the anomalous BFM inputs were arbitrarily mapped to one of the 20 trained position categories, as detailed in Table~\ref{tab:ood_distribution}, despite the absence of any spatial correspondence. Such unreliable misclassifications highlight the system's inability to intrinsically recognize OOD inputs, reinforcing the critical need for a dedicated AD module.

To address this, we integrated ADBlock and first evaluated the impact of the detection threshold $\lambda$ on its performance. As illustrated in Fig.~\ref{fig:detection_metrics}, the TPR increases monotonically as $\lambda$ decreases, indicating that a lower threshold enhances the system's sensitivity to anomalous samples. However, this increased sensitivity comes at the cost of a higher FPR. The F1-score exhibits a distinct peak, identifying the optimal trade-off between anomaly detection capability and false alarm suppression. Thus, this peak value serves as the criterion for selecting the deployment threshold $\lambda$.

To further validate the effectiveness of the proposed method, we compare ADBlock against two established unsupervised AD baselines: OCSVM and iForest. For a fair comparison, we swept the decision threshold for each method to generate their respective TPR-FPR curves. Fig.~\ref{fig:AD_compare} presents the results, with a specific focus on the low-FPR regime ($<10\%$), which is crucial for minimizing false alarms in practical deployments. In this regime, ADBlock achieves a more favorable trade off. Specifically, it attains a TPR of 99.02\% at an FPR of just 0.245\%, while maintaining near perfect detection performance. In contrast, both OCSVM and iForest exhibit reduced efficacy under strict low false alarm constraints. This behavior is expected, as OCSVM relies on learning a nonlinear decision boundary in a kernel induced feature space, which can be sensitive to feature scaling and hyperparameter selection~\cite{scholkopf1999support}. iForest, by comparison, identifies anomalies through random partitioning rather than explicit modeling of the normal data distribution, which may cause low density but still normal BFM fingerprints to be isolated and falsely detected as anomalies~\cite{xu2023deep}. Overall, ADBlock benefits from a reconstruction error criterion that directly measures the conformity of a test sample to the learned manifold of normal BFM patterns.

\begin{figure}[t]
    \centering
    \setlength{\abovecaptionskip}{-2mm}
    \includegraphics[width=1.0\linewidth,trim=20 216 40 258, clip]
    {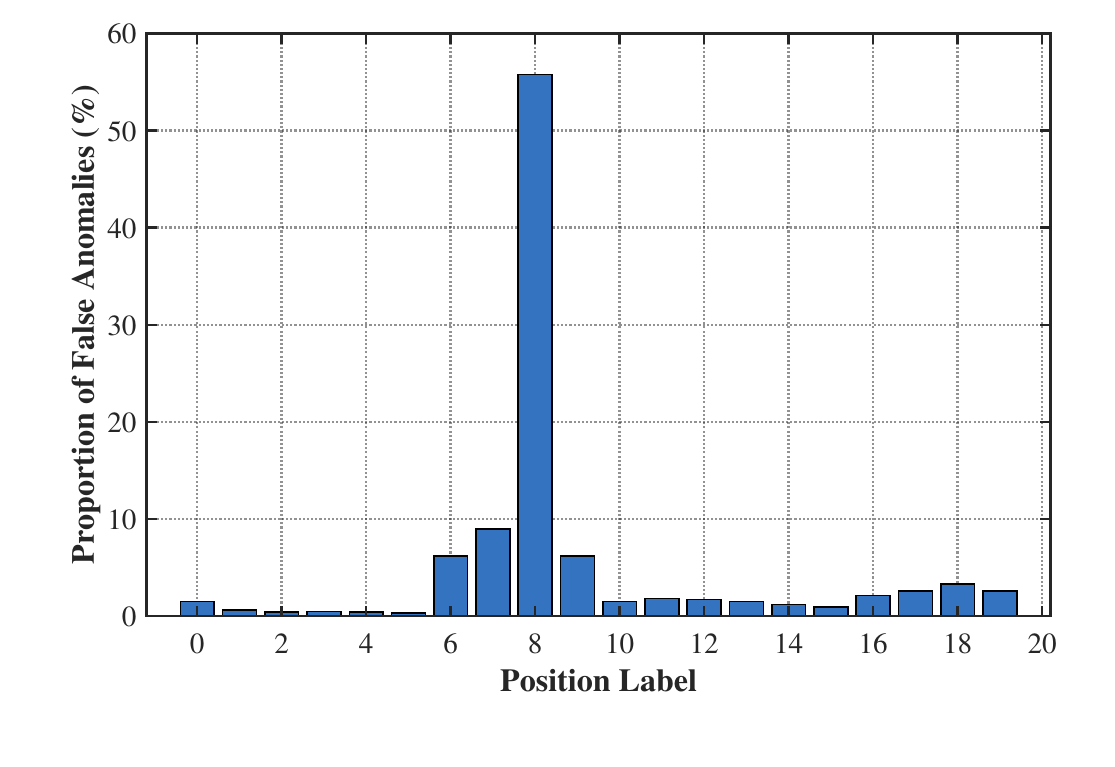} 
    \caption{Distribution of false anomaly counts across different positions. The graph shows a significant spike at position 8, while other positions exhibit a low count of false anomalies.} 
    \label{pos_8}
\end{figure}

\begin{figure}[t]
    \centering
    \setlength{\abovecaptionskip}{5mm}
    \includegraphics[width=0.90\linewidth,trim=30 30 30 20, clip]
    {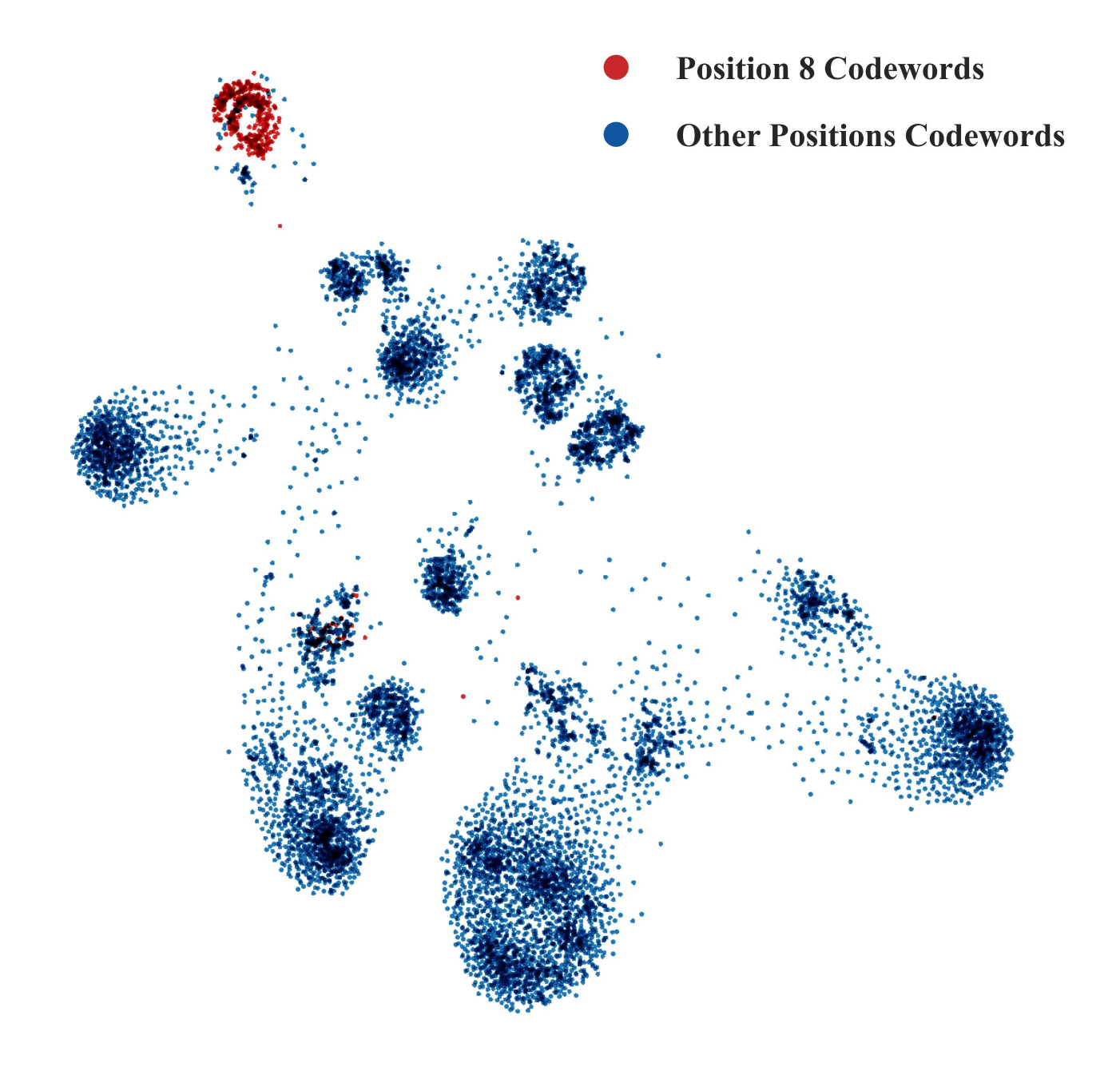}
    \caption{PCA-based visualization of BFM codewords from 20 data collection points. The plot shows the compressed BFM codewords after dimensionality reduction via PCA. Points from position 8 (highlighted in red) exhibit a distribution distinct from the others, indicating a higher tendency for false AD at this position. The remaining points (in blue) form separate clusters, illustrating their separation in the reduced feature space.}
    \label{pos_8_ood}
    \vspace{-12pt}
\end{figure}

We also analyze the distribution of false alarms across positions. Fig.~\ref{pos_8} shows a significant spike in false anomalies at position 8 compared to other points. To understand why, we applied principal component analysis (PCA) to the compressed BFM codewords from all 20 positions (Fig.~\ref{pos_8_ood}). The codewords from position 8 (highlighted in red) form a distinct cluster, separate from the clusters of other positions (in blue). This distributional difference likely causes the higher false alarm rate at position 8, suggesting that location-specific BFM characteristics may require tailored AD thresholds.

Overall, ADBlock substantially enhances AD with minimal false positives. By selecting an appropriate threshold $\lambda$, the system can maintain a high TPR while keeping the FPR low, ensuring robust positioning even when encountering noisy or unexpected BFM data.

\subsubsection{Complexity and Running Time on FPGA}

\begin{table}[t]
  \centering
  \caption{Latency of different neural networks on ``zcu102\_single'' provided by the Deep Learning HDL Toolbox.}
  \label{tab:latency_zcu102_single}
  \begin{tabular}{lccc}
    \toprule
    Model & STA (ms) & AP (ms) & Total (ms) \\
    \midrule
    EFNet           & 0.043 & 0.849 & 0.892 \\
    S-FPNet         & 0.040 & 0.623 & 0.663 \\
    FPNet           & 0.040 & 0.551 & 0.591 \\
    ADBlock & - & 0.316 & 0.316 \\
    \bottomrule
  \end{tabular}
\end{table}

Table \ref{table:adblock} shows the FLOPs of ADBlock, a metric commonly used to quantify the computational complexity of neural networks. However, FLOPs do not account for data loading, storage, scheduling, or parallel execution during hardware deployment. For instance, the latency of a fully connected layer, which can only be executed sequentially, would be significantly higher than that of a convolutional layer capable of high parallelization, even if the latter has higher FLOPs\cite{Wu2019FBNet}. Therefore, to provide a more realistic estimation of the proposed network's complexity, we employ Deep Learning HDL Toolbox of MATLAB
\footnote{MathWorks documentation:\url{https://ww2.mathworks.cn/help/deep-learning-hdl/ug/csi-feedback-with-autoencoders-fpga-implementation.html}} to simulate hardware inference latency. Using the pre-compiled bitstream designed for the Xilinx Zyng UltraScale+ ZCU102 development board with single-precision foating-point arithmetic(``zcu102\_single'' provided by the Deep Learning HDL Toolbox) and setting the hardware clock frequency to 220 MHz, the estimated latencies for each method are summarized in Table \ref{tab:latency_zcu102_single}. Specifically, EFNet requires a total latency of 0.892\,ms (0.043\,ms on the STA and 0.849\,ms on the AP), S-FPNet requires a total latency of 0.663\,ms (0.040\,ms on the STA and 0.623\,ms on the AP), whereas FPNet reduces the end-to-end latency to 0.591\,ms owing to its more parallelizable architecture, i.e., a 10.9\% reduction relative to S-FPNet. The \emph{ADBlock} runs solely on the AP with an additional 0.316\,ms; when combined with FPNet, the overall latency becomes 0.907\,ms.

\section{Conclusion}

In this paper, we presented FPNet, a DL-based framework designed to jointly optimize BFM feedback compression and indoor positioning. The proposed framework efficiently addresses the challenge of reducing communication overhead in Wi-Fi systems while maintaining high positioning accuracy. Experimental results show that FPNet outperforms traditional protocol-based methods, achieving up to 97.52\% positioning accuracy with only 100 feedback bits and delivering significant improvements in both BFM feedback reconstruction and positioning performance. To further assess robustness in more complex and dynamic environments, we conducted additional experiments in a dynamic office setting with changed human activity and multipath conditions, and showed that FPNet can rapidly recover its performance via fine-tuning when the deployment environment drifts after training.
Furthermore, we introduced ADBlock, an anomaly-detection module that enhances FPNet's robustness by identifying OOD BFM samples. The inclusion of ADBlock ensures that the system remains reliable in realistic deployments, where users may be located outside predefined spatial regions or environmental conditions may change. Our experimental evaluation indicates that ADBlock achieves 99\% detection accuracy with a false alarm rate below 1.5\%, successfully detecting anomalies with minimal impact on positioning performance. Overall, these results highlight the potential of DL-based solutions for jointly optimizing BFM feedback and positioning tasks, and underscore the importance of incorporating anomaly-detection mechanisms to ensure system resilience.

Some challenges are worth exploring further in the future.
One important direction is to improve scalability to the larger antenna arrays and denser OFDM grids of Wi-Fi~6/7 and beyond, where both the input dimension of FPNet and the associated feedback overhead increase significantly. This calls for more structured architectures that explicitly exploit joint spatial and frequency correlations, such as tensorized encoders, factorized or low-rank decoders, and attention mechanisms
~\cite{Guo2020CsiNetPlus,Suto2021RadioMapDL}.
Besides, as future Wi-Fi systems expand across multiple frequency bands, frequency-adaptive of FPNet will be needed. Approaches such as shared latent spaces with band-specific adapters, or meta-learning and domain-adaptation techniques~\cite{Suto2021RadioMapDL,Zinys2021DomainIndependentGAN}, may enable rapid retargeting of the model across heterogeneous multi-band environments, enhancing FPNet’s applicability in future Wi-Fi sensing and communication systems.

\bibliographystyle{IEEEtran}
\bibliography{reference}

\end{document}